\documentclass[prb,preprint,superscriptaddress,showpacs,footinbib]{revtex4}

\usepackage{amsmath, amsfonts, amssymb}
\usepackage{graphicx}

\begin{document}

\title{Ultrafast screening and carrier dynamics in ZnO: theory and experiment}

\author{Marijn A.M. Versteegh}
\affiliation{Debye Institute for Nanomaterials Science, Utrecht University, Princetonplein 1, 3584 CC Utrecht, The Netherlands}
\author{Tim Kuis}
\affiliation{Debye Institute for Nanomaterials Science, Utrecht University, Princetonplein 1, 3584 CC Utrecht, The Netherlands}
\author{H.T.C. Stoof}
\affiliation{Institute for Theoretical Physics, Utrecht University, Leuvenlaan 4, 3584 CE Utrecht, The Netherlands}
\author{Jaap I. Dijkhuis\footnote{j.i.dijkhuis@uu.nl}}
\affiliation{Debye Institute for Nanomaterials Science, Utrecht University, Princetonplein 1, 3584 CC Utrecht, The Netherlands}

\begin{abstract}
At carrier densities above the Mott density Coulomb screening destroys the exciton resonance. This, together with band-gap renormalization and band filling, severely affects the optical spectra. We have experimentally studied these effects by ultrafast pump-probe reflectivity measurements on a ZnO single crystal at various wavelengths around the exciton resonance and in a broad carrier-density range. Theoretically we determined the Mott density in ZnO to be $1.5\times10^{24}$ m$^{-3}$ at 300 K. Taking a field-theoretical approach, we derived and solved the Bethe-Salpeter ladder equation and we computed the density-dependent reflectivity and absorption spectra. A carrier dynamics model has been developed, containing three-photon absorption, carrier cooling, and carrier trapping near the surface. The agreement between the theoretical reflectivity based on our model and the experimental data is excellent.
\end{abstract}


\maketitle

\section{Introduction}

Because of its wide direct band gap of 3.37 eV, ZnO has many possible applications for optoelectronic devices, including solar cells and light-emitting diodes. ZnO nanowires are used as waveguides and UV lasers,\cite{huang 2001,johnson 2001,johnson 2003,van vugt 2006} photodetectors, \cite{soci 2007} and optical switches. \cite{kind 2002} For such applications, it is important to know and understand the optical spectra of ZnO at high carrier densities, as well as the carrier dynamics.

The optical spectra at high carrier densities are strongly influenced by screening, band-gap renormalization (BGR), and band filling. At densities higher than the so-called Mott density $n_{M}$, screening of the Coulomb interaction destroys the exciton resonance. Here we present a concise theoretical and experimental study of these phenomena in ZnO, covering the exciton regime, the electron-hole plasma regime, and the crossover between them. Analysis of the pump-probe reflectivity experiment described here also reveals the ultrafast carrier dynamics near the crystal surface.

Pump-probe reflectivity experiments on ZnO by other groups \cite{c. li 2005, schneck 2008, magoulakis 2010} have shown large reflectivity changes at high carrier densities. Despite these striking observations, a quantitative picture of the carrier dynamics and the reflectivity spectrum at high densities does not exist. This has several causes: (1) A thorough understanding of the physics of a high-density electron-hole gas in ZnO is lacking. (2) Around the exciton resonance at 3.31 eV, no pump-probe reflectivity measurements have yet been reported. (3) All experiments were carried out at very high carrier densities, far above $n_M$. The ultrafast carrier and reflectivity dynamics in the exciton regime and across the crossover from the exciton regime to the electron-hole-plasma (EHP) regime have not yet been studied. (4) For a straightforward theoretical analysis of pump-probe reflectivity data one needs a homogeneous carrier density within the penetration depth of the reflected probe. It is the aim of the present paper to report in considerable detail on progress in all these four directions. This is achieved in the following manner.

In Secs. \ref{experimental method} and \ref{experimental results} of this paper, we present pump-probe reflectivity data on a ZnO single crystal, taken at four probe wavelengths around the exciton resonance. We used 800-nm pump pulses to ensure a homogeneous carrier density within the penetration depth of the reflected probe. Excitation took place via three-photon absorption (3PA). Measurements were performed in a broad density range of $10^{22}-10^{26}$ m$^{-3}$, to probe the dynamics both above and below the Mott density.

The experimental data are compared with theory. In Sec. \ref{theory} we compute the Mott density and the electron-hole chemical potential. Using the solutions of the statically screened Bethe-Salpeter equation we then compute the density-dependent optical spectra. In Sec. \ref{ultrafast carrier dynamics}, by comparing our theoretical results with the experiment, conclusions will be drawn about the ultrafast carrier dynamics. Finally in Sec. \ref{z-scan}, the obtained intensity-dependent 3PA coefficient is tested by a Z-scan measurement.

\section{\label{experimental method}Pump-probe method}

\begin{figure*}
\includegraphics[width=\textwidth]{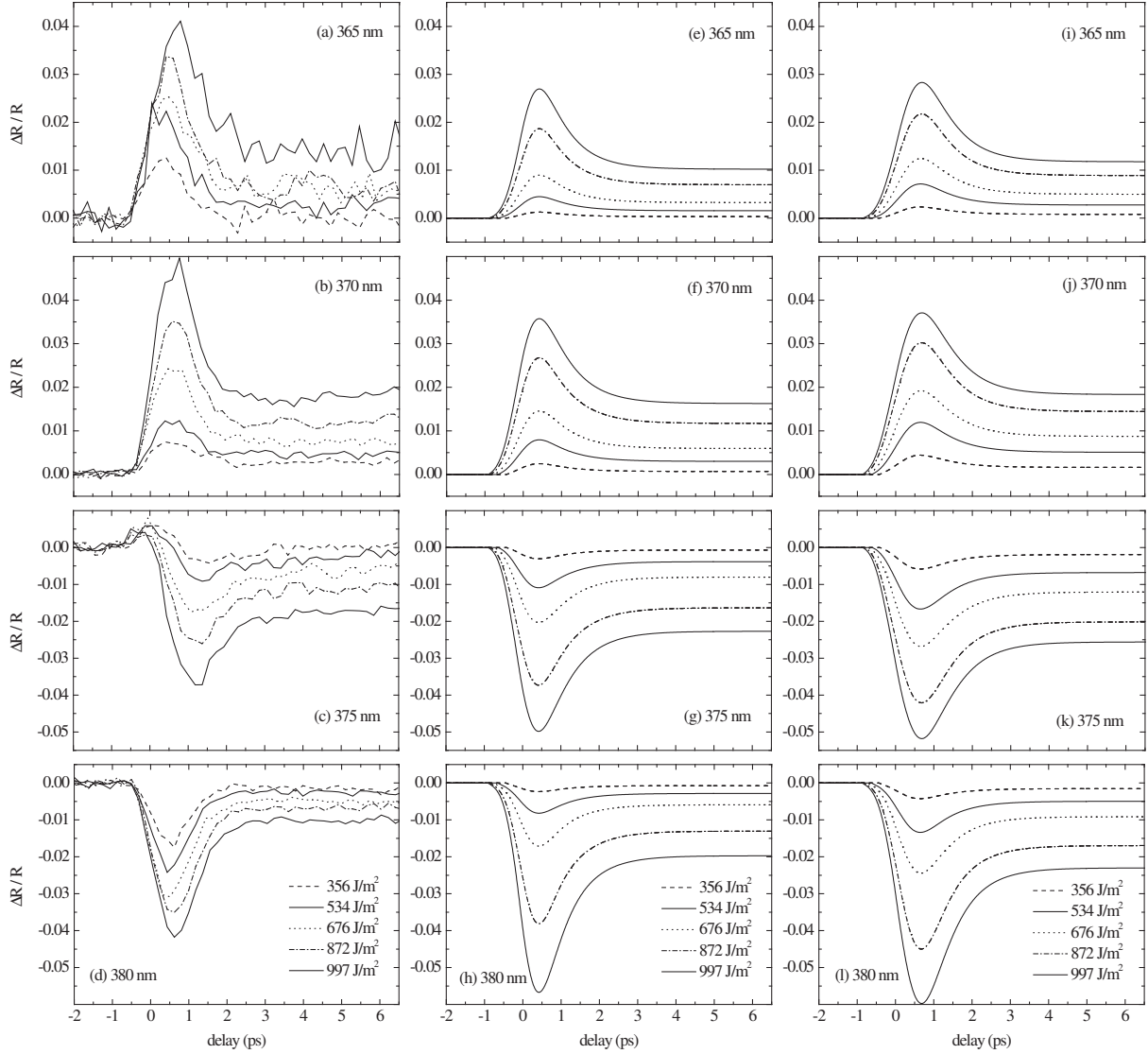}
\caption{Dynamics of the reflectivity (reflection coefficient / reflectance) of a ZnO single crystal surface following a 1.41-ps 800-nm pump pulse. (a-d) Pump-probe measurement results at probe wavelengths of (a) 365 nm (3.397 eV), (b) 370 nm (3.351 eV), (c) 375 nm (3.306 eV), and (d) 380 nm (3.263 eV). At all probe wavelengths we took the same fluence series. (e-h) Fits according to the Simple Model [Eq. (\ref{Simple Model})]. (i-l) Fits according to the Saturation and Cooling Model [Eq. (\ref{Sophisticated Model})].\label{fig1}}
\end{figure*}

\begin{figure*}
\includegraphics[width=\textwidth]{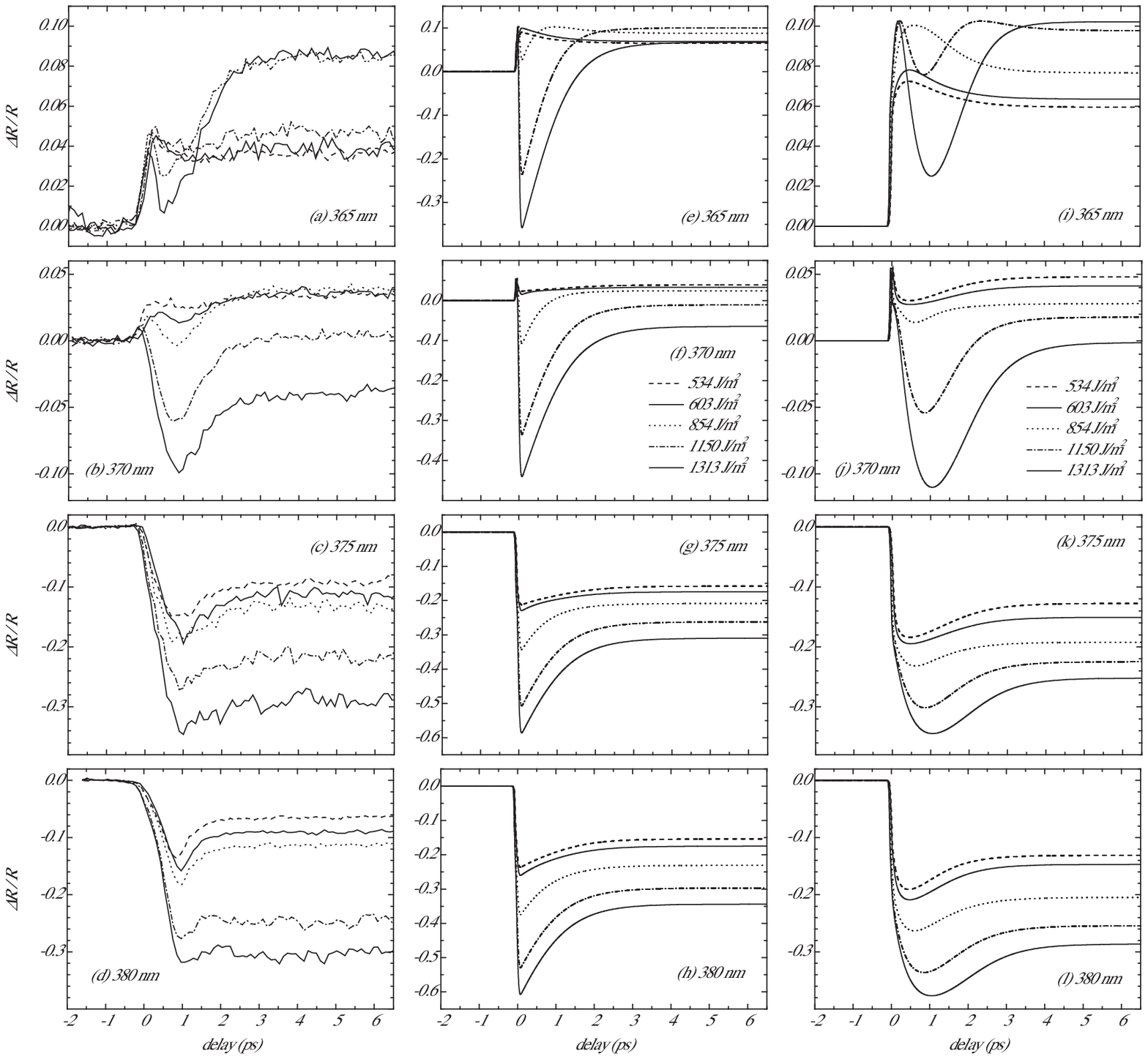}
\caption{Dynamics of the reflectivity of a ZnO single crystal surface following a 141-fs 800-nm pump pulse. (a-d) Pump-probe measurement results. Fluences at all probe wavelengths according to the legend of (f) and (j). (e-h) Fits according to the Simple Model [Eq. (\ref{Simple Model})]. Note the different vertical scale here. (i-l) Fits according to the Saturation and Cooling Model [Eq. (\ref{Sophisticated Model})].\label{fig2}}
\end{figure*}

For the experiment, 800-nm laser pulses from an amplified 1-kHz Ti:sapphire laser were split into a pump pulse and a probe pulse. The pump pulse was sent through a 500-Hz chopper wheel and a delay line, and then focused into the sample. The probe pulse was focused into a 4.5-mm thick sapphire crystal for self-focusing and white-light generation. The beam was subsequently sent through a BBO crystal for sum-frequency generation of 800-nm light and a selected frequency from the white-light pulse. Undesired wavelengths were filtered out. Then the probe was sent through a polarization rotator to obtain \textit{s}-polarization and focused onto the center of the pump spot on the front surface of the sample. By changing the orientation of the BBO crystal the probe was tuned to any desired wavelength between 360 and 440 nm, with a spectral resolution of 2 nm (FWHM).

As sample we used an epi-polished ZnO single crystal ($5\times5\times0.523$ mm$^3$), purchased from MTI Corp. It is oriented in the [0001] direction, i.e., with the \textit{c}-axis perpendicular to the plane of the wafer. The electric field of the probe was polarized perpendicularly to the \textit{c}-axis. Experiments were performed on the Zn face of the crystal. The angle of incidence was 0$^{\circ}$ for the pump and 22.3$^{\circ}$ for the probe. The pump spot on the sample was 220 $\mu$m in diameter (FWHM), the probe 35 $\mu$m. The probe pulses reflected at the front surface of the sample were detected by a photodiode and a lock-in amplifier. Measurements were performed with ($1.41 \pm 0.10$)-ps and ($141 \pm 5$)-fs pump pulses (FWHM), and with 365-nm, 370-nm, 375-nm, and 380-nm probe wavelengths, at room temperature.

\section{\label{experimental results}Pump-probe results}

\begin{figure}
\includegraphics[width=0.5\textwidth]{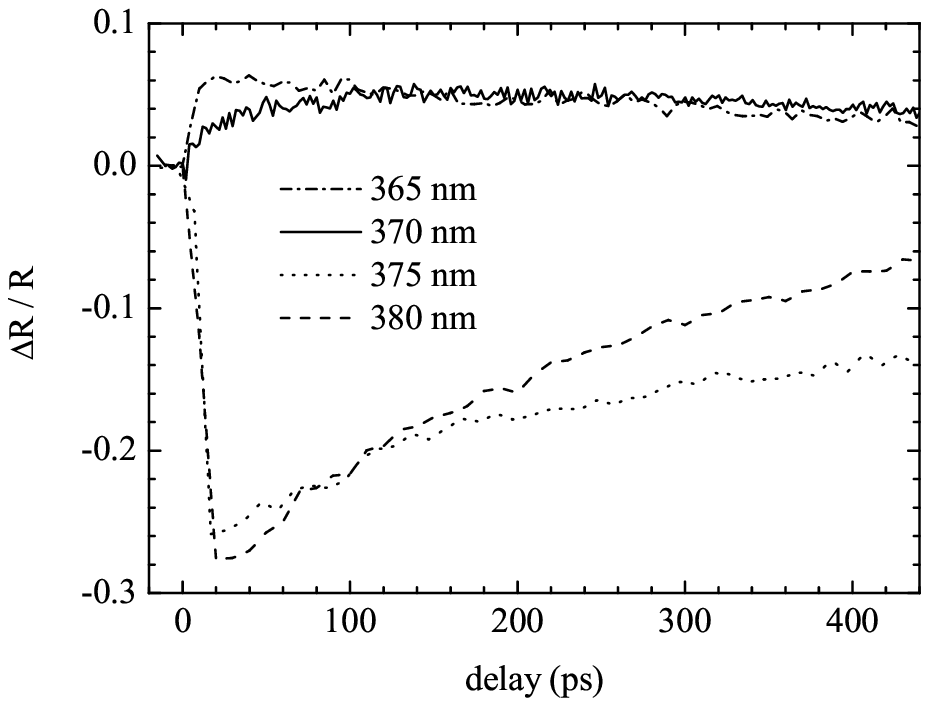}
\caption{Pump-probe reflectivity results with a long delay. Pump: 141-fs 800-nm pulses. Fluences: 1150 J/m$^2$ for the 365-nm measurement and 1313 J/m$^2$ for the 370-nm, 375-nm, and 380-nm measurements.\label{fig3}}
\end{figure}

The results of the pump-probe measurements with the 1.41-ps pulses are shown in Fig. \ref{fig1} (a-d). We observe at 365 and 370 nm that the reflectivity increases while at 375 and 380 nm it decreases with increasing pump fluence. For all measurements each extremum is followed by a fast relaxation to a plateau different from the initial level.

Figure \ref{fig2} (a-d) shows the results of the measurements with 141-fs pump pulses. Here the carrier densities reached are much higher because of the nonlinearity of the absorption. At 375 and 380 nm we see again a decrease of the reflectivity during 1 ps. At the highest pump fluence $\Delta R/R=-0.3$. At 365 and 370 nm at the highest fluences the pump-probe signal increases, decreases, and finally increases to a plateau.

Pump-probe reflectivity results with long delays are shown in Fig. \ref{fig3}. We observe that the decay from the plateau to the original reflectivity takes several hundreds of picoseconds.

When one encounters complicated pump-probe traces as the 1313 J/m$^2$-result in Fig. \ref{fig2}(a), one might be inclined to search for independent processes that explain the first sharp peak, the subsequent dip, and the rise again to the plateau.

Actually, as we will demonstrate in this paper, these rich phenomena can be simply explained by increase and subsequent decrease of carrier density, at least if we correctly account for the highly nonlinear relation between reflectivity and carrier density. In Sec. \ref{theory} we theoretically compute the optical spectra for several carrier densities and in particular this relation. In Sec. \ref{ultrafast carrier dynamics} the theoretical results of Sec. \ref{theory} will be combined with a model for the carrier dynamics. The reader who is exclusively interested in the results for the optical spectra and the carrier dynamics can simply first consider Figs. \ref{fig9} and \ref{fig11} and then proceed to Sec. \ref{ultrafast carrier dynamics}.

\section{\label{theory}Theory}

In order to elucidate how the reflectivity changes with increasing carrier density one has to study how the complex index of refraction changes with increasing carrier density. Li \textit{et al.} \cite{c. li 2005} compared their pump-probe results with a free-carrier Drude model, described in Ref. \onlinecite{sokolowskitinten 2000}. This model is correct for very high carrier densities where the Coulomb interaction between the carriers is almost completely screened. At carrier densities about $10^{28}$ m$^{-3}$ this is indeed the case. However, at our carrier densities of $10^{22}-10^{26}$ m$^{-3}$, and at our probe wavelengths, the Coulomb attraction between electrons and holes does play a major role in the optical properties. Indeed, below the Mott density it produces the exciton resonance in the absorption and reflectivity spectra. As carrier density increases, screening gradually destroys the exciton resonance.

Next to the vanishing of the exciton resonance due to screening, band-gap renormalization determines the optical properties: the band gap shrinks for increasing carrier density due to exchange and correlation effects. A final important effect for the optical spectra is band filling. We first quantitatively address these three phenomena and subsequently compute the den\-sity-depen\-dent absorption and reflectivity spectra.

ZnO has one conduction band and three valence bands, called A, B, and C. Each of these four bands is twofold degenerate because of the spin degree of freedom. The valence bands are split by the crystal field and the spin-orbit coupling: the AB splitting equals 10 meV, the AC splitting 44 meV.\cite{lambrecht 2002} At present it is unclear how the band-gap renormalization behaves in case of multiple split valence bands. Our goal is to set up a simple description of the many-body physics to explain our experimental data. For that purpose we make throughout this paper the simplification to take only the conduction band and the A valence band into account. Although we realize that this will affect our results quantitatively in certain parameter regimes of the experiment, we do not expect it to affect the physics of interest to us qualitatively. Transitions between the conduction band and the A valence band are allowed without spin-flip for the probe polarization in our experiment $\mathbf{E}\perp c$.\cite{thomas 1960, lambrecht 2002} In this two-band model there is a single band gap of $E_{G,0}=3.372$ eV.\cite{jellison 1998} Further, we use isotropic parabolic bands and quasi-equilibrium of the electron-hole gas, so that equilibrium statistical mechanics can be used to describe its properties.

\subsection{Coulomb screening and Mott density}\label{4a}

In this section we consider the screening of the Cou\-lomb interaction and compute the Mott density $n_{M}$. The Mott density marks the crossover between the density regime where excitons exist (the exciton regime) and the density regime where they are screened away (the EHP regime). It is important to pin down this value, not only for understanding the optical properties of ZnO, but also to know whether certain observed phenomena in ZnO, such as lasing, have an excitonic nature, as is frequently claimed, or not. The published values for the room-temperature Mott density in ZnO largely vary \cite{johnson 2003, klingshirn 2007, klingshirn book 2007, dai 2010, chen 2001, sun 2005, ozgur 2005, arai 2006, djurisic 2006} and range from $3\times10^{23}$ to $3.7\times10^{25}$ m$^{-3}$.

The physics of unscreened excitons is equal to that of hydrogen atoms. The ground-state binding energy is related to the Bohr radius $a_{0}$ by
\begin{equation}
E_{0}=\frac{\hbar^{2}}{2m_{r}a_{0}^{2}},
\end{equation}
where $m_{r}=(1/m_{e}+1/m_{h})^{-1}$ is the reduced mass of the electron-hole pair. The electron mass in the conduction band and the hole mass in the A valence band have been experimentally determined to be $m_{e}=0.28 m_{0}$ (Ref. \onlinecite{button 1972}) and $m_{h}=0.59 m_{0}$ (Ref. \onlinecite{hummer 1973}), respectively, so that $m_{r}=0.19 m_{0}$. Here $m_{0}$ denotes the bare electron mass. The exciton binding energy is known to be 60 meV, from which it follows, in agreement with literature, \cite{klingshirn 2007} that $a_{0}=1.83$ nm. The Bohr radius also obeys the relation
\begin{equation}
a_{0}=\frac{4\pi\hbar^{2}\varepsilon_{r}\varepsilon_{0}}{e^{2}m_{r}},
\end{equation}
from which we extract the relative dielectric constant $\varepsilon_{r}=6.56$. Note that we use SI units throughout this paper.

We describe the screened Coulomb interaction by the Yukawa potential
\begin{equation}\label{yukawa}
V_{s}(\mathbf{x}-\mathbf{x'})=\frac{e^{2}}{4\pi\varepsilon_{0}\varepsilon_{r}|\mathbf{x}-\mathbf{x'}|}e^{-|\mathbf{x}-\mathbf{x'}|/\lambda_{s}},
\end{equation}
where $\lambda_{s}$ is the screening length. The derivation of the Yukawa potential needs the approximation of static screening, \cite{haug koch 2004} that is, screening is established fast with respect to the Fermi frequencies of the charge carriers. This is a good approximation if $\hbar$ times the plasma frequency
\begin{equation}
\omega_{p}=\sqrt{\frac{e^{2}n}{\varepsilon_{0}\varepsilon_{r}m_{r}}},
\end{equation}
is high with respect to the Fermi energies of the electrons and holes
\begin{equation}\label{Fermi energy}
\varepsilon_{F,i}=\frac{\hbar^{2}}{2m_{i}}(3\pi^{2}n)^{2/3},
\end{equation}
where $i$ stands for $e$ (electron) or $h$ (hole). The conditions $\hbar\omega_{p}>\varepsilon_{F,e}$ and $\hbar\omega_{p}>\varepsilon_{F,h}$ are both met if $n<2.8\cdot10^{26}$ m$^{-3}$. Since in our experiment the carrier density does not exceed this value, we can use the Yukawa potential [Eq. (\ref{yukawa})]. Note that we always consider the situation that the electron density is equal to the hole density, $n_{e}=n_{h}$. This is necessarily true for optical excitation. This density we call the carrier density $n$.

In an electron plasma the screening length is given by \cite{haug koch 2004}
\begin{equation}\label{screening length}
\lambda_{s,e}=\sqrt{\frac{\varepsilon_{0}\varepsilon_{r}}{e^{2}}\frac{\partial \mu_{e}}{\partial n}},
\end{equation}
where $\mu_{e}$ is the chemical potential. We have electrons \textit{and} holes however. In an electron-hole plasma the screening length is related to the screening lengths of the electron and hole plasmas according to
\begin{equation}\label{electron hole screening}
\lambda_{s}^{-2}=\lambda_{s,e}^{-2}+\lambda_{s,h}^{-2}.
\end{equation}

For the Fermi-Dirac distribution at zero temperature, Eq. (\ref{screening length}) reduces to the Thomas-Fermi screening length. In the classical high-temperature limit, the particles have a Boltzmann distribution and Eq. (\ref{screening length}) reduces to the Debye-H\"{u}ckel screening length. \cite{haug koch 2004} We do not take any of these limits, but we compute $\lambda_{s,e}$ and $\lambda_{s,h}$ using numerically determined ideal-gas chemical potentials, calculated from
\begin{equation}\label{chemical potential ideal gas}
n=\frac{1}{2\pi^{2}}(\frac{2m_{i}}{\hbar^{2}})^{3/2}\int_{0}^{\infty}\textrm{d}\varepsilon\sqrt{\varepsilon}\frac{1}{e^{\beta(\varepsilon-\mu_{i})}+1},
\end{equation}
where $\beta=1/(k_{B}T)$. The electron chemical potential $\mu_{e}$ is measured from the conduction band edge, the hole chemical potential $\mu_{h}$ from the valence band edge.

\begin{figure}
\includegraphics[width=0.5\textwidth]{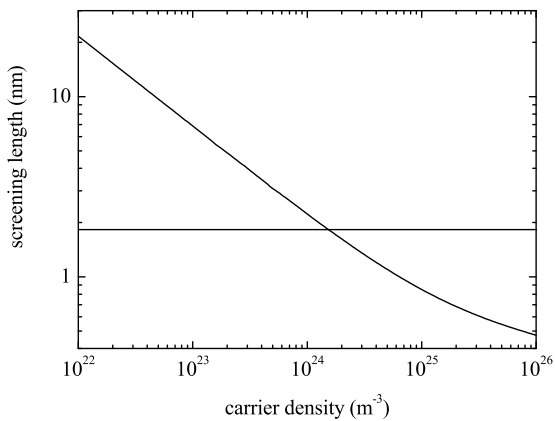}
\caption{Screening length $\lambda_{s}$ at 300 K versus carrier density. Note the logarithmic scales. The horizontal line indicates the exciton Bohr radius $a_{0}$. The Mott density $n_{M}$, i.e. the density at which $\lambda_{s}=a_{0}$, is equal to $1.5\times10^{24}$ m$^{-3}$. Excitons only exist at densities below $n_{M}$.\label{fig4}}
\end{figure}

Of course, the electron-hole gas is not an ideal gas. Coulomb interactions play a very important role. That is why we compute the screening length. The Coulomb interactions affect the chemical potentials via the possible presence of excitons, but in order to be able to compute the chemical potentials including the Coulomb interactions, one must first know the screening length. For the computation of the screening length the ideal gas approximation turns out to be sufficient, as will be demonstrated in Sec. \ref{4b}. Note that BGR does not affect the screening length at a certain carrier density. The result for $\lambda_{s}$ at 300 K is given in Fig. \ref{fig4}.

If one approximates the Yukawa potential by the Hulth\'{e}n potential, which deviates only very little from the Yukawa potential, one can analytically compute the ground-state exciton binding energy in the case of screening as \cite{banyai 1986, haug koch 2004}
\begin{equation}\label{screened exciton binding energy}
E_{s}=\left\{ \begin{array}{ll}
(1-a_{0}/\lambda_{s})^{2}E_{0}\quad & \text{if \ \  }\lambda_{s}\geq a_{0},\\
0 & \text{if \ \  }\lambda_{s}<a_{0}.
\end{array} \right.
\end{equation}
The Mott density $n_M$ is the carrier density at which $\lambda_{s}=a_{0}$. At this density $E_{s}=0$ and excitons can no longer exist. Our calculation shows that $n_{M}=1.5\times10^{24}$ m$^{-3}$. This value is lower than the values given in literature, except for the values given by Klingshirn \textit{et al.}\footnote{Most authors base their value of the Mott density on a derivation in Ref. \onlinecite{klingshirn book 1995} (pp. 306-307) within Debye-H\"{u}ckel screening theory. Debye-H\"{u}ckel screening theory gives a reasonable approximation for the screening length at room temperature for densities around and below $10^{24}$ m$^3$. However, the derivation in Ref. \onlinecite{klingshirn book 1995} needs three comments: (1) The equation for the Debye-H\"{u}ckel screening length misses a factor $1/\sqrt{4\pi}$. This has been corrected in the next edition.\cite{klingshirn book 2007} (2) In the equation relating the exciton Bohr radius to the exciton binding energy a factor $1/(4\pi)$ is missing. (3) Only the screening from one type of carriers has been taken into account. The semiconductor however contains electrons \emph{and} holes. The last two errors lead to a Mott density of $3.7\times10^{25}$ m$^{-3}$, a factor $8\pi$ too large.}

Apart from screening of the Coulomb attraction between electrons and holes there is of course also screening of the Coulomb repulsion between carriers of the same kind. The effects of Coulomb repulsion and the screening of it on the energy levels are captured in the band-gap renormalization.

\subsection{Chemical potential}\label{4b}

The electron-hole pair chemical potential with respect to the band gap $\mu=\mu_{e}+\mu_{h}$ describes band filling and is an important parameter for the optical properties. A positive $\mu$ means population inversion. In this section we will no longer use the ideal gas model of Eq. (\ref{chemical potential ideal gas}), but a more accurate model involving Coulomb interactions. In this interaction model we take into account that below the Mott density part of the electrons and holes are bound into excitons.

The unbound electrons and holes obey Fermi-Dirac statistics  and have the distribution functions
\begin{equation}\label{Fermi-Dirac distribution}
f_{i}(\varepsilon)=\frac{1}{e^{\beta(\varepsilon-\mu_{i})}+1}.
\end{equation}
Excitons, however, obey Bose-Einstein statistics:
\begin{equation}\label{Bose-Einstein distribution}
f_{ex}(\varepsilon)=\frac{1}{e^{\beta(\varepsilon-\mu)}-1}.
\end{equation}
The exciton chemical potential is the electron-hole pair chemical potential $\mu$. The energy of the exciton is its kinetic energy minus the binding energy. We only consider excitons in the ground state, so $\varepsilon=\varepsilon_{kin}-E_{s}$ for the excitons. In this model we suppose that if an exciton's kinetic energy is higher than its binding energy, it immediately dissociates.

There are four possible spin states of the exciton,
\begin{equation*}
|s,m_{s}\rangle\ \in\ \{\ |0,0\rangle,\ |1,-1\rangle,\ |1,0\rangle,\ |1,1\rangle\ \}.
\end{equation*}
Hence the density of states of excitons is twice as large as that of electrons or holes. Only excitons in the states $|0,0\rangle$ and $|1,0\rangle$ can be created by a photon and can recombine into a photon without a spin-flip. In the computation of the susceptibility therefore exclusively the states $|0,0\rangle$ and $|1,0\rangle$ are to be taken into account. In the computation of the chemical potential, however, all four states have to be taken into account, since they all four contribute to the density of states at equilibrium. This gives the following relation between the exciton density and the exciton chemical potential
\begin{equation}\label{exciton bose gas}
n_{ex}=\frac{1}{\pi^{2}}(\frac{2(m_{e}+m_{h})}{\hbar^{2}})^{3/2}\int_{-E_{s}}^{0}\textrm{d}\varepsilon\sqrt{\varepsilon+E_{s}}f_{ex}(\varepsilon),
\end{equation}
with $f_{ex}(\varepsilon)$ given by Eq. (\ref{Bose-Einstein distribution}). Since the electron density is equal to the hole density, the electron and hole chemical potentials can be calculated from the following system of two equations with two unknowns:
\begin{equation}\label{excitonic fraction}
n=n_{ex}+\frac{1}{2\pi^{2}}(\frac{2m_{i}}{\hbar^{2}})^{3/2}\int_{0}^{\infty}\textrm{d}\varepsilon\sqrt{\varepsilon}f_{i}(\varepsilon),
\end{equation}
for $i=e$ and $i=h$ and with $n_{ex}$ given by Eq. (\ref{exciton bose gas}) and $f_{i}(\varepsilon)$ given by Eq. (\ref{Fermi-Dirac distribution}).

\begin{figure}
\includegraphics[width=0.5\textwidth]{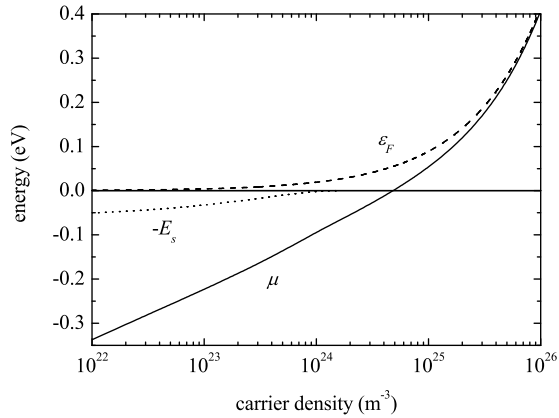}
\caption{The Fermi energy $\varepsilon_{F}=\varepsilon_{F,e}+\varepsilon_{F,h}$ [Eq. (\ref{Fermi energy})], the electron-hole pair chemical potential $\mu$, and the exciton ground energy level $-E_{s}$ at 300 K [Eq. (\ref{screened exciton binding energy})], versus carrier density.\label{fig5}}
\end{figure}

The result at 300 K is displayed in Fig. \ref{fig5}. Also $-E_{s}$ and $\varepsilon_{F}=\varepsilon_{F,e}+\varepsilon_{F,h}$ are shown in this figure. For carrier densities larger than $4.8\times10^{24}$ m$^{-3}$ the chemical potential is positive (population inversion) and for increasing density it approaches the Fermi energy. The exciton binding energy decreases with carrier density due to screening of the Coulomb attraction and becomes zero at the Mott density.

\begin{figure}
\includegraphics[width=0.5\textwidth]{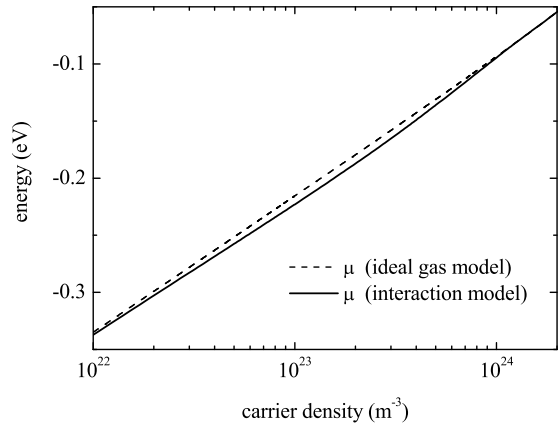}
\caption{The pair chemical potential $\mu$ at 300 K according to the ideal gas model and the interaction model.\label{fig6}}
\end{figure}

In Fig. \ref{fig6} the chemical potential in the interaction model is compared with the chemical potential in the ideal gas model. We find that at 300 K there is little difference, confirming that our calculation of the screening length in the previous section is a good approximation.

\subsection{Excitonic fraction}

\begin{figure}
\includegraphics[width=0.5\textwidth]{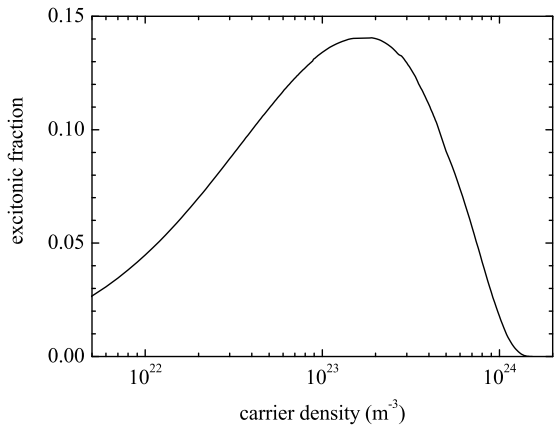}
\caption{Fraction of carriers that are bound into excitons at 300 K versus carrier density.\label{fig7}}
\end{figure}

In principle the theory in Sec. \ref{4a} overestimates the screening, since Eqs. (\ref{screening length}) and (\ref{electron hole screening}) apply to a plasma of unbound carriers. Below the Mott density a fraction of the carriers is bound into excitons, and screening by excitons is weaker than screening by unbound carriers.

In order to study how large this effect on the screening is, we compute the excitonic fraction $n_{ex}/n$ from Eqs. (\ref{excitonic fraction}). The result is shown in Fig. \ref{fig7}. The maximum of the excitonic fraction is 0.14 at $n=1.9\times10^{23}$ m$^{-3}$. The vast majority of carriers is thus not bound at room temperature.

To estimate the error made in our calculation of the screening, we make the rather extreme assumption that the excitons do not contribute to screening at all. Then at $n=1.9\times10^{23}$ m$^{-3}$, where the excitonic fraction is the highest, only a density of $1.63\times10^{23}$ m$^{-3}$ contributes to screening. Instead of a screening length of 5.00 nm we find $\lambda_{s}=5.35$ nm and the excitonic fraction becomes 0.156. If we repeat the calculation with this new excitonic fraction, we get $\lambda_{s}=5.40$ nm and an excitonic fraction of 0.159. We conclude that the errors in the screening length as a result of using an ideal EHP theory are at most about 8\%.

For the rest of this paper we use the screening length from ideal EHP theory, as given in Fig. \ref{fig4}, and the chemical potential according to the interaction model, as given in Figs. \ref{fig5} and \ref{fig6}.

\subsection{Band-gap renormalization}

\begin{figure}
\includegraphics[width=0.5\textwidth]{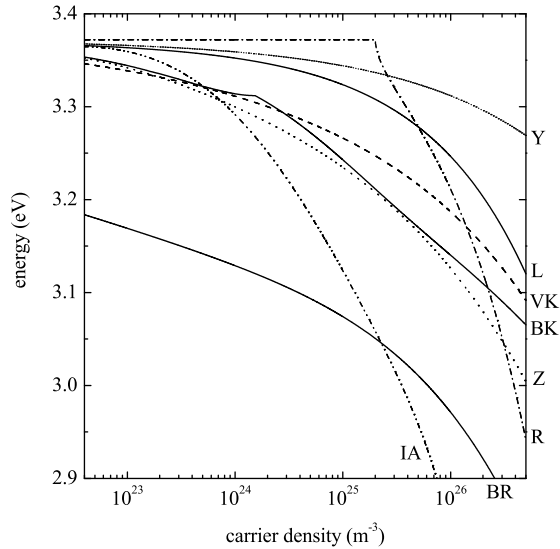}
\caption{Band gap renormalization: the band gap of ZnO versus carrier density. BR: Beni and Rice \cite{beni 1978} with numerical results taken from Ref. \onlinecite{yamamoto 2002}; VK: Vashishta and Kalia \cite{vashishta 1982}; R: Roth \textit{et al.} \cite{roth 1982}; BK: B\'{a}nyai and Koch \cite{banyai 1986}; Z: Zimmermann \cite{zimmermann 1988}; IA: Inagaki and Aihara \cite{inagaki 2002}; Y: Ye \textit{et al.} \cite{ye 2005}; L: Lu \textit{et al.} \cite{lu 2007}\label{fig8}}
\end{figure}

Klingshirn \textit{et al.} \cite{klingshirn 2007} took a phenomenological approach to relate the band-gap renormalization to the Coulomb screening. From experiment it is known that, within errors, the exciton resonance does not shift with increasing carrier density. Our measurements reported in this paper confirm this observation. From this result one must conclude that the increase by screening of the exciton resonance energy is compensated by the decrease due to BGR. This means that the density at which the BGR equals $E_{0}=60$ meV, and hence $E_{G}=3.312$ eV, is the Mott density.

This is a valid argument, but the problem remains that it is notoriously difficult to determine, experimentally or theoretically, the BGR in ZnO. In the literature very different results for the BGR have been reported. Without claiming to be complete, we show several of them in Fig. \ref{fig8}. Where applicable we inserted $E_{0}=60$ meV, $a_{0}=1.828$ nm, $T=300$ K, and $E_{G,0}=3.372$ eV.\footnote{Note that the exponent in the numerator in Eq. (3) of Ref. \onlinecite{zimmermann 1988} should be $1/2$ instead of $1/4$.} All of these BGR results, except one, could in principle be used for the determination of $n_{M}$. Only the BGR graph of B\'{a}nyai and Koch cannot be used, because the use of their formula requires that one already knows the magnitude of screening. Different values for the BGR lead to different values for the Mott density. Therefore we believe that our method for calculating $n_{M}$ is more reliable.

In our calculations of the optical properties we use the phenomenological BGR formula of B\'{a}nyai and Koch, \cite{banyai 1986, haug koch 2004} according to which BGR exactly compensates the effect of screening on the position of the exciton resonance:
\begin{equation}\label{BGR}
E_{G}=\left\{ \begin{array}{ll}
E_{G,0}-E_{0}+E_{s}\quad &\text{if \ \ }n\leq n_{M},\\
E_{G,0}-E_{0}a_{0}/\lambda_{s}\quad &\text{if \ \ }n>n_{M}.
\end{array} \right.
\end{equation}
$E_{s}$ is given by Eq. (\ref{screened exciton binding energy}).

\subsection{\label{susceptibility section}Susceptibility}

In the Appendix the Bethe-Salpeter equation is derived from quantum field theory and solved in the statically screened ladder approximation. It is derived that the susceptibility of an electron-hole gas in a direct semiconductor is given by
\begin{equation}\label{BS susceptibility}
\chi_{BS}(\omega)=\frac{2d_{cv}}{\varepsilon_{0}L^{3}}\sum_{\mathbf{k}}\chi_{\mathbf{k}}(\omega).
\end{equation}
Here, $d_{cv}$ is the dipole moment of the valence-conduction band transition, $L^{3}$ is the volume of the (cubic) crystal, and BS stands for `Bethe-Salpeter'. The summation over all $\mathbf{k}$-states of the carriers is three-dimensional with a step size of $2\pi/L$ in each direction. A factor 2 is included to account for the spin degree of freedom. The auxiliary function $\chi_{\mathbf{k}}(\omega)$ is implicitly given by the susceptibility integral equation
\begin{equation}\label{chik}
\chi_{\mathbf{k}}(\omega)=\chi_{k}^{0}(\omega)\Big(1+\frac{1}{d_{cv}L^{3}}\sum_{\mathbf{k}'}V_{s,|\mathbf{k}-\mathbf{k'}|}\chi_{\mathbf{k}'}(\omega)\Big).
\end{equation}
Here, the mean-field function $\chi_{k}^{0}(\omega)$, depending only on the length of $\mathbf{k}$, is given by
\begin{equation}\label{chik0}
\chi_{k}^{0}(\omega)=-d_{cv}\frac{1-f_{k,e}-f_{k,h}}{\hbar(\omega+i\gamma(\omega))-\varepsilon_{k,e}-\varepsilon_{k,h}-E_{G}},
\end{equation}
where $\varepsilon_{k,i}=\hbar^{2}k^{2}/(2m_{i})$ are the kinetic energies of the electrons and holes, $f_{k,i}=f_{i}(\varepsilon)$ are the Fermi-Dirac distribution functions, and $\gamma(\omega)$ is the frequency-dependent damping, which we discuss in more detail below.

Finally, $V_{s,|\mathbf{k}-\mathbf{k'}|}$ is the Yukawa potential in momentum space, i.e. the Fourier transform of Eq. (\ref{yukawa}),
\begin{equation}
V_{s,|\mathbf{k}-\mathbf{k'}|}=\frac{e^{2}}{\varepsilon_{0}\varepsilon_{r}}\;\frac{1}{k^{2}+k'^{2}-2kk'\cos\theta+\lambda_{s}^{-2}},
\end{equation}
where $\theta$ is the angle between $\mathbf{k}$ and $\mathbf{k'}$. As convention for the Fourier transform of the potential we use
\begin{equation}\label{Fourier potential}
V_s(\mathbf{x})=\frac{1}{L^{3}}\sum_{\mathbf{k}}V_{s,\mathbf{k}}e^{i\mathbf{k}\cdot\mathbf{x}}\:\:\text{and}\quad V_{s,\mathbf{k}}=\!\int \!\textrm{d}\mathbf{x} V_{s}(\mathbf{x})e^{-i\mathbf{k}\cdot\mathbf{x}}.
\end{equation}
We note that it is also possible to derive Eq. (\ref{BS susceptibility}) using an equations-of-motions approach.\cite{haug koch 2004}

We solve Eq. (\ref{chik}) by using a matrix inversion method, described by Haug and Koch.\cite{haug koch 2004} Because of rotation symmetry we can replace $V_{s,|\mathbf{k}-\mathbf{k'}|}$ by its angle-averaged
\begin{equation}
\begin{split}
\overline{V}_{s,k,k'}&=\frac{1}{2}\int_{0}^{\pi}\textrm{d}\theta\; \sin\theta\; V_{s,|\mathbf{k}-\mathbf{k'}|}\\
&=\frac{e^{2}}{4\varepsilon_{0}\varepsilon_{r}kk'}\ln\Big[\frac{(k^{2}+k'^{2}+2kk')\lambda_{s}^{2}+1}{(k^{2}+k'^{2}-2kk')\lambda_{s}^{2}+1}\Big],
\end{split}
\end{equation}
and transform the three-dimensional summation over $\mathbf{k'}$ into a one-dimensional summation over its length $k'$,
\begin{equation}\label{chik1d}
\chi_{k}(\omega)\!=\!\chi_{k}^{0}(\omega)\Big[1+\frac{s}{d_{cv}(2\pi)^{3}}\!\!\!\sum_{k'=0,s,\ldots}\!\!\!\!\!4\pi k'^{2}\,\overline{V}_{s,k,k'}\,\chi_{k'}(\omega)\Big].
\end{equation}
Here $s$ is the step size of the resulting $k$-summation.\footnote{Note that in Ref. \onlinecite{haug koch 2004} the factor $s/(2\pi)^3$ from the new step size, and the factor $4\pi k'^{2}$ from the integration over the angles, are not explicitly shown.}

We introduce the vertex function $\Gamma_{k}(\omega)$ as
\begin{equation}\label{def gamma}
\chi_{k}(\omega)=\Gamma_{k}(\omega)\chi_{k}^{0}(\omega).
\end{equation}
Inserting this into Eq. (\ref{chik1d}), we obtain the integral equation
\begin{equation}
\Gamma_{k}(\omega)=1+\frac{s}{2\pi^{2}d_{cv}} \sum_{k'=0,s,\ldots} \!\!\!\! k'^{2}\;\overline{V}_{s,k,k'}\;\chi_{k'}^{0}(\omega)\,\Gamma_{k'}(\omega).
\end{equation}
When $\Gamma_{k}$ is seen as a vector, this is an equation of the form
\begin{equation}
\overrightarrow{\Gamma}(\omega)=\overrightarrow{1}+\overrightarrow{\overrightarrow{\mathrm{M}}}(\omega)\cdot\overrightarrow{\Gamma}(\omega),
\end{equation}
where $\overrightarrow{1}$ is the unit vector and $\overrightarrow{\overrightarrow{\mathrm{M}}}(\omega)$ is the matrix
\begin{equation}
\mathrm{M}_{k,k'}(\omega)=\frac{s}{2\pi^{2} d_{cv}} k'^{2}\;\overline{V}_{s,k,k'}\;\chi_{k'}^{0}(\omega).
\end{equation}
We see that
\begin{equation}\label{matrix inversion}
\overrightarrow{\Gamma}(\omega)=[\overrightarrow{\overrightarrow{1}}-\overrightarrow{\overrightarrow{\mathrm{M}}}(\omega)]^{-1}\cdot\overrightarrow{1},
\end{equation}
where $\overrightarrow{\overrightarrow{1}}$ is the unit matrix.

Also a background susceptibility $\chi_{L}$ produced by the lattice, including the valence electrons, should be included. This yields the following expression for the susceptibility of the semiconductor:
\begin{equation}\label{susceptibility}
\chi(\omega)=\chi_{L}+\frac{2d_{cv}s}{(2\pi)^{3}\varepsilon_{0}}\sum_{k=0,s,\ldots}^{k_{max}}\!\!\!\!4\pi k^{2}\chi_{k}(\omega).
\end{equation}
Here we again performed a transformation from the three-dimensional $\mathbf{k}$-sum to the one-dimensional $k$-sum and we added an upper limit.

\subsection{Optical spectra}

The complex refractive index $\tilde{n}(\omega)=n'(\omega)+i n''(\omega)$ is related to the complex susceptibility $\chi(\omega)=\chi'(\omega)+i\chi''(\omega)$ as $\tilde{n}(\omega)=\sqrt{1+\chi(\omega)}$. The reflectivity (reflection coefficient) $R(\omega)$ of $s$-polarized light (the probe light in our experiment) is related to the complex index of refraction \cite[p. 422]{pedrotti 1993} and reads
\begin{equation}\label{reflectivity}
R=1-\frac{4a\cos i}{\cos^{2}i+2a\cos i+\sqrt{b^{2}+4n'^{2}n''^{2}}},
\end{equation}
with $a=(b^{2}+4n'^{2}n''^{2})^{1/4}\cos[\frac{1}{2}\arctan(2n'n''/b)]$, $b=n'^{2}-n''^{2}-\sin^{2}i$ and $i$ the angle of incidence. The absorption coefficient is given by
\begin{equation}\label{absorption coefficient}
\alpha(\omega)=\alpha_{I}+\frac{2\omega n''(\omega)}{c},
\end{equation}
where $c$ is the vacuum speed of light and $\alpha_{I}$ is absorption due to crystal impurities, relatively very small and frequency-independent within the frequency range of our experiment.

In order to find the density-dependent absorption and reflectivity spectra, we thus compute for each carrier density (1) $\chi_{k}^{0}(\omega)$ from Eq. (\ref{chik0}), and (2) the inverse of the matrix $\overrightarrow{\overrightarrow{1}}-\overrightarrow{\overrightarrow{\mathrm{M}}}(\omega)$, inserting appropriate values for the chemical potential, screening length, damping and band gap renormalization. The spectra can then be obtained via Eqs. (\ref{matrix inversion}), (\ref{def gamma}), and (\ref{susceptibility}-\ref{absorption coefficient}).

\begin{figure*}
\includegraphics[width=\textwidth]{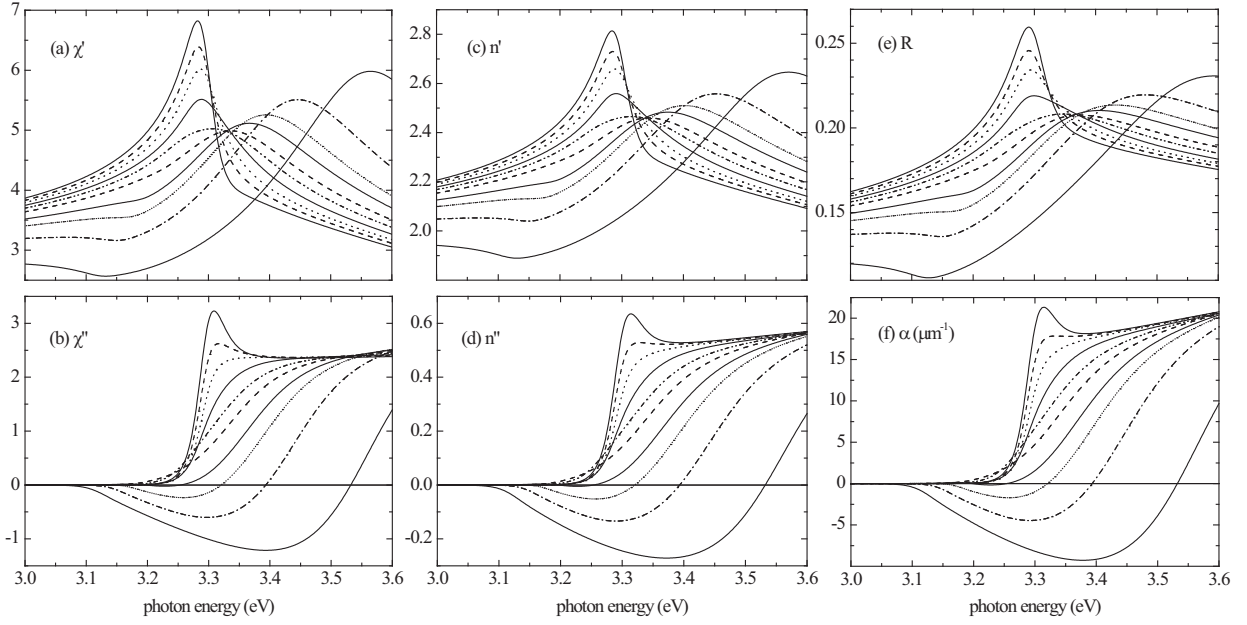}
\caption{Theoretical spectra of excited ZnO at 300 K for $\mathbf{E}\perp c$: (a) real part of the susceptibility, (b) imaginary part of the susceptibility, (c) real part of the index of refraction, (d) imaginary part of the index of refraction, (e) reflectivity at $i=22.3^{\circ}$ [Eq. (\ref{reflectivity})], (f) absorption coefficient [Eq. (\ref{absorption coefficient})]. The carrier densities in all graphs are, from the highest curve to the lowest curve: $5\times10^{21}$ (solid), $5\times10^{23}$ (dash), $1\times10^{24}$ (dot), $2\times10^{24}$ (solid), $5\times10^{24}$ (dash dot dot), $1\times10^{25}$ (dash), $2\times10^{25}$ (solid), $3\times10^{25}$ (short dot), $5\times10^{25}$ (dash dot), and $1\times10^{26}$ m$^{-3}$ (solid). In all graphs the gradual disappearance of the exciton resonance due to screening is visible.\label{fig9}}
\end{figure*}

For the computation of $\chi_{k}^{0}(\omega)$, a damping function $\gamma(\omega)$ is needed, representing the combined effect of carrier-phonon, carrier-impurity, and carrier-carrier scattering. In Ref. \onlinecite{khitrova 1999} a microscopic description of the damping due to carrier-carrier scattering is given. In order to stay close to the experiment, however, we here choose to work with a phenomenological function $\gamma(\omega)$. A frequency-dependence is necessary to correctly describe the Urbach tail, i.e. the decrease of absorption on the long-wavelength side of the exciton resonance.\cite{haug koch 2004} In our calculation we take
\begin{equation}\label{damping}
\gamma(\omega)=\frac{\gamma_{0}}{e^{(-\hbar\omega+E_{G}-E_{s}-E_{\alpha})/E_{\alpha}}+1}.
\end{equation}
This function gives the best agreement between the theoretical results for the optical spectra at the lowest carrier densities and measured optical spectra of unexcited ZnO. \cite{bond 1965, yoshikawa 1997, jellison 1998, muth 1999} In principle, at higher densities damping is stronger because of increased carrier-carrier scattering. We choose however to work with a density-independent damping in order to reduce the number of parameters.

For the numerical computation we choose a step size $s=5\times10^7$ m$^{-1}$ and an upper limit $k_{max}=2.5\times10^9$ m$^{-1}$. As a result our to be inverted matrix has a size of $51\times51$. We checked that with smaller step sizes the same results are obtained, but with a longer computation time. With a larger step size one obtains unphysical fluctuations in the spectra.

The results of the computation for the complex susceptibility, the complex refractive index, the reflectivity at $i=22.3^{\circ}$, and the absorption coefficient are presented in Fig. \ref{fig9}. All low-density spectra exhibit an exciton peak. The exciton peak in the absorption spectrum [Fig. \ref{fig9}(f)] is at 3.31 eV, precisely where it should, a first evidence that our theory works well. The exciton resonance disappears from the spectra at densities around the Mott density. This is a second support for our results, or reversely, for the value of the Mott density that we obtained earlier. For densities exceeding $2\times10^{25}$ m$^{-3}$ negative absorption, i.e. gain, appears.

\begin{figure}
\includegraphics[width=0.5\textwidth]{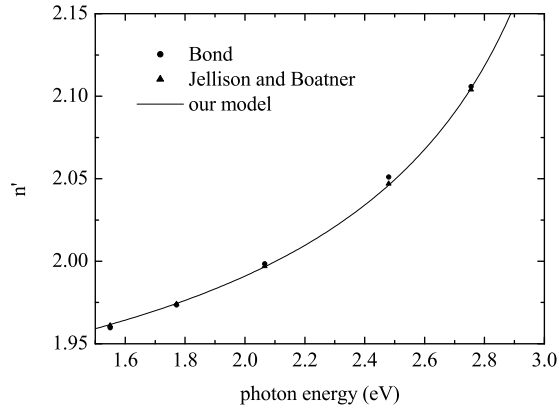}
\caption{Real part of the refractive index of unexcited ZnO. Our model is compared with the long-wavelength experimental data of Bond \cite{bond 1965} and Jellison and Boatner. \cite{jellison 1998}\label{fig10}}
\end{figure}

Our theory has in principle six free parameters. Table \ref{table 1} shows their values. We have determined these values by fitting the low-density result of our theory to experimental data on unexcited ZnO. Our theory therefore has predictive power for higher carrier densities. Five parameters were determined by fitting the theoretical spectra at low density to the published experimental data \cite{bond 1965, yoshikawa 1997, jellison 1998, muth 1999} on the linear absorption and refractive index spectra of unexcited ZnO, both near and far from the exciton resonance. Our value for $\gamma_{0}$ is higher than the room temperature damping parameters of Refs. \onlinecite{jellison 1998} and \onlinecite{hauschild 2006}, but lower than that of Ref. \onlinecite{yoshikawa 1997}. In Fig. \ref{fig10} the real refractive index from our model is compared with the long-wavelength experimental data of Refs. \onlinecite{bond 1965} and \onlinecite{jellison 1998}. The impurity absorption coefficient $\alpha_{I}$ was determined by a simple measurement of the transmission through our 523 $\mu$m thick ZnO crystal at wavelengths around 400 nm. This measurement shows that $\alpha_{I}=1.1\times10^3$ m$^{-1}$, a factor $\sim10^4$ smaller than the absorption coefficients near the band gap.

\begin{table}
\caption{Model parameters.}\label{table 1}
\begin{ruledtabular}
\begin{tabular}{cc}
Parameter  & Value\\
\hline
$d_{cv}$ & $4.2\times10^{-29}$ Cm\\
$k_{max}$ & $2.5\times10^9$ m$^{-1}$\\
$\chi_{L}$ & 2.4\\
$\hbar\gamma_{0}$ & 50 meV\\
$E_{\alpha}$& 22 meV\\
$\alpha_{I}$& $1.1\times10^3$ m$^{-1}$
\end{tabular}
\end{ruledtabular}
\end{table}

\begin{figure}
\includegraphics[width=0.5\textwidth]{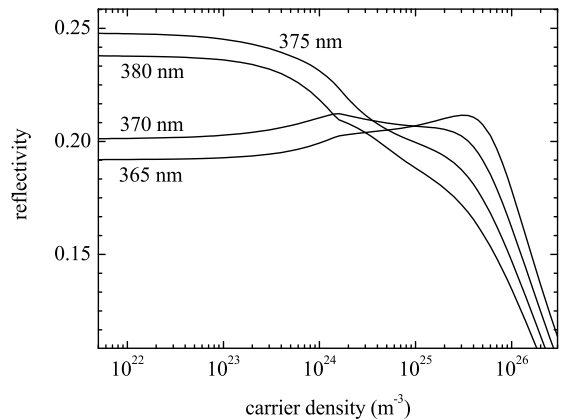}
\caption{Theoretical reflectivity for $i=22.3^{\circ}$ at 365 nm (3.397 eV), 370 nm (3.351 eV), 375 nm (3.306 eV), and 380 nm (3.263 eV), versus carrier density.\label{fig11}}
\end{figure}

In order to connect our pump-probe measurement results to theory, we calculate the reflectivity versus density at 365, 370, 375, and 380 nm. The results are shown in Fig. \ref{fig11}. The kink at the Mott density in some of the graphs is due to the kink in the BGR formula used [Eq. (\ref{BGR})].

We see at 365 and 370 nm that reflectivity rises with increasing carrier density, reaches a maximum, and decreases again. This behavior can be related to the pump-probe signals of Figs. \ref{fig1}(a,b) and \ref{fig2}(a,b). For 1.41-ps pump pulses the carrier density remains left of this maximum, while for 141-fs pulses it goes beyond the top until maximum carrier density is reached at the bottom of the dip in the pump-probe signal. After that point, the density decreases again.

At 375 and 380 nm reflectivity monotonically decreases with increasing density. This is consistent with all experimental data at those wavelengths, except for a tiny peak in Fig. \ref{fig1}(c).

\section{\label{ultrafast carrier dynamics}Ultrafast carrier dynamics}

Our extensive experimental results on the reflectivity versus time (Figs. \ref{fig1}(a-d), \ref{fig2}(a-d), and \ref{fig3}), combined with our theoretical results on the reflectivity versus carrier density (Fig. \ref{fig11}), allow for the determination of the carrier density versus time. As we will show now, a concise carrier dynamics model can be found accounting for all measurements, performed at different probe wavelengths, both above and below the exciton resonance, at several fluences, both for long pump pulses (low density) and short pump pulses (high density).

\subsection{Buildup and decay}

Since the band gap (3.37 eV) and the exciton energy (3.31 eV) are larger than two times the photon energy (1.55 eV), absorption of an 800-nm pulse is a three-photon process. 3PA of 800-nm pulses in ZnO has been reported by He \textit{et al.} \cite{he 2005} and Dai \textit{et al.} \cite{dai 2005}. Thanks to the large penetration depth, 3PA provides a homogeneous carrier density over the penetration depth of the reflected probe (about 50 nm). This presents clear advantage of 3PA over one-photon absorption.

Following carrier buildup, we observe at all probe wavelengths, both for 1.41-ps and 141-fs pulses, a fast relaxation to a reflectivity level higher or lower than the initial reflectivity [Figs. \ref{fig1}(a-d) and \ref{fig2}(a-d)]. The subsequent decay to the initial level takes hundreds of picoseconds (Fig. \ref{fig3}), in agreement with decay times measured in time-resolved photoluminescence experiments.\cite{reynolds 2000, koida 2003, guo 2003, bauer 2004, johnson 2004, wilkinson 2004, teke 2004} This slow decay is the result of radiative and nonradiative recombination of carriers and excitons. The remainder of this paper is devoted to extracting the carrier dynamics during the first 6 ps after the pump pulse. On this timescale the slow decay can be safely ignored.

In line with literature, \cite{magoulakis 2010, bauer 2004, shi 2008} we explain the fast decay by trapping of carriers into impurities, such as oxygen vacancies. It is known that the density of singly ionized oxygen vacancy traps in a surface layer of 30-100 nm is much higher than in the interior of the crystal. \cite{vanheusden 1996} The fast decay therefore mainly occurs in this surface layer. This idea is supported by the 30-nm thick surface-recombination layer found by Shalish \textit{et al.} \cite{shalish 2004} and with the observation of Magoulakis \textit{et al.} \cite{magoulakis 2010} that increased surface roughness leads to higher trapping efficiencies. To explain the relaxation to the plateau, we make a distinction between charge carriers near the surface (the surface carriers) and charge carriers in the interior of the crystal (the bulk carriers). Only the surface carriers are subject to fast decay. After about 2 ps all surface carriers have been trapped and the remaining bulk carriers produce the reflectivity plateau.

Alternative explanations for the fast decay, like Auger recombination and stimulated emission, can be excluded, since at low pump fluences and low densities the fast decay is as prominent in the pump-probe results as at high fluences and high densities.

Trap saturation cannot explain the reflectivity plateaus. Such an explanation requires a single limited trap density. From the pump-probe results of Figs. \ref{fig1}(a-d) and \ref{fig2}(a-d), however, it can be found, using Fig. \ref{fig11} as a gauge, that for 1.41-ps pump pulses the plateau is reached after a fast density decay in the order of $10^{23}$ m$^{-3}$, while for 141-fs pulses it is reached after a decay in the order of $10^{25}$ m$^{-3}$.

\subsection{Simple Model}

\begin{figure*}
\includegraphics[width=0.9\textwidth]{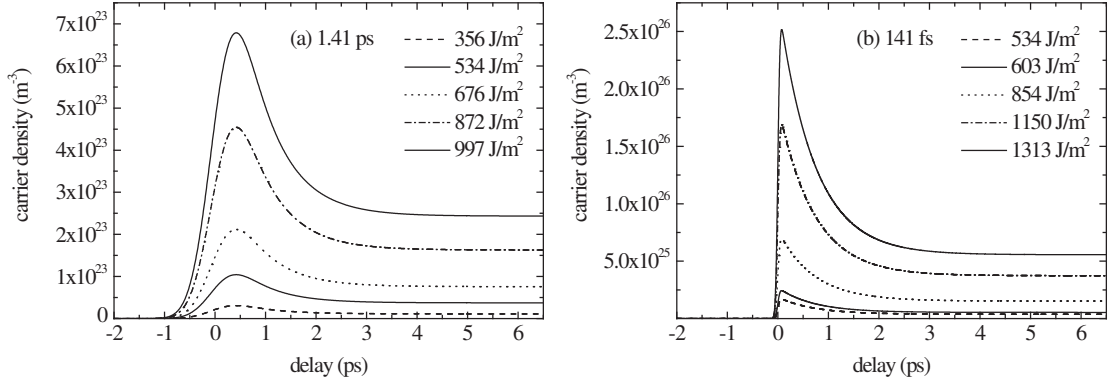}
\caption{Carrier density versus time deduced from experiment using the Simple Model for (a) 1.41-ps and (b) 141-fs 800-nm pump pulses.\label{fig12}}
\end{figure*}

The above considerations form the basis of a simple carrier dynamics model, the `Simple Model'. The carrier density is taken to be homogeneous within the penetration depth of the reflected probe. This density $n$, being the sum of the surface carrier density within the probe penetration depth $n_{S}$ and the bulk carrier density within the probe penetration depth $n_{B}$, $n=n_{S}+n_{B}$, determines the reflectivity according to Fig. \ref{fig11}. We do not take spatial variations in the refractive index into account. In this model the carriers are in thermal equilibrium at $T=300$ K.

The mathematical expression for the Simple Model is
\begin{equation}\label{Simple Model}
\begin{split}
\frac{\textrm{d} n_{S}(t)}{\textrm{d}t}&=\frac{S\alpha_{3} I(t)^3}{3\hbar\omega}-\frac{n_{S}(t)}{\tau},\\
\frac{\textrm{d} n_{B}(t)}{\textrm{d}t}&=\frac{(1-S)\alpha_{3} I(t)^3}{3\hbar\omega},
\end{split}
\end{equation}
with the initial conditions $n_{S}(-\infty)=n_{S}(-\infty)=0$. Here $\alpha_{3}$ is the 3PA coefficient for 800-nm light, $\hbar\omega=1.55$ eV is the pump photon energy, $\tau$ is the surface carrier decay time, $S$ is the fraction of the carriers within the probe penetration depth that are near the surface, and $I(t)$ is the intensity of the pump pulse
\begin{equation}\label{intensity}
I(t)=\frac{[1-R(\omega)]f}{\sqrt{2\pi}d}e^{-t^2/(2d^2)},
\end{equation}
where $R(\omega)=0.105$ is the reflectivity of the 800-nm pump, $f$ is the pump fluence, and $d$ measures the pulse length: $141/\sqrt{8\ln2}=60$ fs or 600 fs.

By fitting the Simple Model to the experimental results, we arrive at $\alpha_{3}=5\times10^{-27}$ m$^{3}$/W$^{2}$, $\tau=0.7$ ps, and $S=0.8$. Our trapping time of 0.7 ps is in good agreement with values reported in literature. \cite{magoulakis 2010, bauer 2004, shi 2008}

Figure \ref{fig12} shows the dynamics of the carrier density $n$ as deduced from the measurements using the Simple Model. Note that for 1.41 ps pump pulses the carrier density remains below the Mott density, while for 141-fs pulses densities in the order of $10^{26}$ m$^{-3}$ are reached.

Combining Fig. \ref{fig12} with Fig. \ref{fig11} yields the theoretical reflectivity versus time, Fig. 1(e-h) and Fig. 2(e-h). The agreement with the experimental results is surprisingly good, both with respect to the shapes of the pump-probe results as with respect to the absolute values of $\Delta R / R$.

Our value for $\alpha_{3}$ is a factor 2 lower than the value of $(1.0\pm0.2)\times10^{-26}$ m$^{3}$/W$^{2}$, reported by He \textit{et al.} \cite{he 2005}, but one must note that their value was obtained for intensities $I<4\times10^{14}$ W/m$^2$, while in our 1.41-ps measurements intensities of $6\times10^{14}$ W/m$^2$ and in our 141-fs measurements intensities of $8\times10^{15}$ W/m$^2$ are reached. At high intensities 3PA gets saturated and $\alpha_{3}$ decreases with increasing intensity. \cite{gu 2008} The more sophisticated `Saturation and Cooling Model' described in the next section takes this effect into account.

\subsection{\label{Saturation and Cooling Model}Saturation and Cooling Model}

\begin{figure*}
\includegraphics[width=0.9\textwidth]{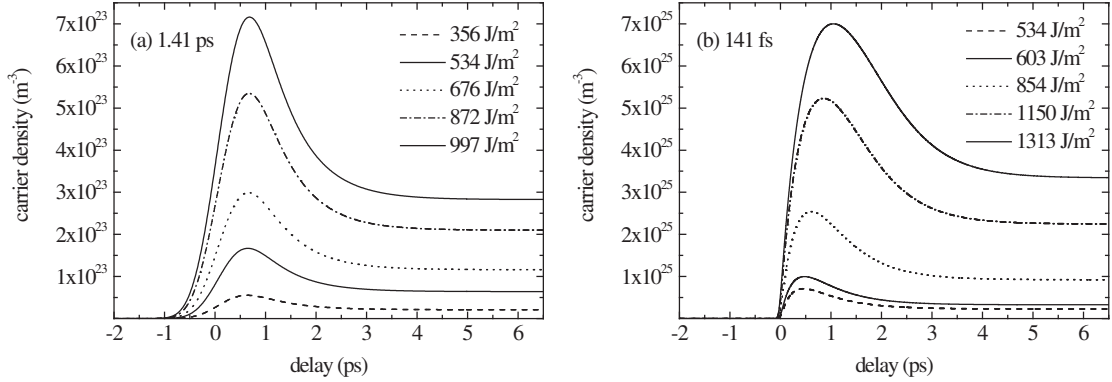}
\caption{Thermalized carrier density versus time deduced from experiment using the Saturation and Cooling Model for (a) 1.41-ps and (b) 141-fs 800-nm pump pulses.\label{fig13}}
\end{figure*}

Comparing the theoretical traces of Fig. \ref{fig2}(e-h) with the experimental ones of Fig. \ref{fig2}(a-d), we see that the plateau levels are in good agreement. However, the negative peaks are too sharp and too large. In Fig. \ref{fig1} the agreement is better, but still there is some room for improvement: the low-fluence peaks are too small. In this section we describe a `Saturation and Cooling Model' that takes 3PA saturation and carrier cooling into account, and matches the experimental results better than the Simple Model described in the previous section. The rate equations read
\begin{equation}\label{Sophisticated Model}
\begin{split}
\frac{\textrm{d} n_{H}(t)}{\textrm{d}t}&=\frac{\alpha_{3}[I(t)] I(t)^3}{3\hbar\omega}-\frac{n_{H}(t)}{k(n_{tot})},\\
\frac{\textrm{d} n_{S}(t)}{\textrm{d}t}&=\frac{S n_{H}(t)}{k(n_{tot})}-\frac{n_{S}(t)}{\tau},\\
\frac{\textrm{d} n_{B}(t)}{\textrm{d}t}&=\frac{(1-S) n_{H}(t)}{k(n_{tot})},
\end{split}
\end{equation}
with initial conditions $n_{H}(-\infty)=n_{S}(-\infty)=n_{B}(-\infty)=0$. The total carrier density $n_{tot}=n_{H}+n_{S}+n_{B}$. Saturation of 3PA is described by an intensity-dependent $\alpha_{3}$. Like in the Simple Model, $\tau=0.7$ ps, $S=0.8$, and the intensity is given by Eq. (\ref{intensity}).

In the Saturation and Cooling Model it is recognized that carriers are created high in the bands and consequently do not immediately contribute to the susceptibility and the optical properties at optical frequencies near the band gap. Furthermore their contribution to screening is negligible because of their high kinetic energy. The optical properties are governed by the thermalized carriers only.

In Eq. (\ref{Sophisticated Model}) $n_{H}$ is the density of carriers high in the bands. They are created via 3PA and cool down to thermalized surface carriers $n_{S}$ and thermalized bulk carriers $n_{B}$ with cooling time $k(n_{tot})$. The reflectivity is, like in the Simple Model, determined by $n=n_{S}+n_{B}$ via Fig. \ref{fig11}.

The resulting carrier density responsible for the optical response $n_{S}+n_{B}$ is given in Fig. \ref{fig13}. The theoretical reflectivity versus time is given in Figs. \ref{fig1}(i-l) and \ref{fig2}(i-l). The sharp peaks of the Simple Model are smoothed and the peaks have amplitudes that are in much better accordance with the experimental results. Note that even the complicated high-fluence results of Fig. \ref{fig2}(a,b) are faithfully described by this model.

Gu \textit{et al.} \cite{gu 2008} have studied 3PA saturation of 780-nm light. They experimentally determined $\alpha_{3}(I)$ to be
\begin{equation}\label{gu}
\alpha_{3}[I]_{Gu}=\frac{\alpha_{3,\textit{Gu}}^{0}}{1+I^3/I_{S}^3},
\end{equation}
with $\alpha_{3,\textit{Gu}}^{0}=1.3\times10^{-26}$ m$^{3}$/W$^{2}$ and $I_{S}=4.4\times10^{14}$ W/m$^2$. In our opinion this equation cannot be correct at high intensities, because according to it the total carrier density does not exceed $4\times10^{23}$ m$^{-3}$, while it is evident that in our experiment at least a factor $10^2$ higher densities are reached. By fitting the Saturation and Cooling Model to our experimental results we have found
\begin{equation}\label{3PA saturation}
\alpha_{3}[I]=\alpha_{3}^{A}+\frac{\alpha_{3}^{B}}{1+I^3/I_{S}^3},
\end{equation}
with $\alpha_{3}^{A}=3\times10^{-27}$ m$^{3}$/W$^{2}$, $\alpha_{3}^{B}=7\times10^{-27}$ m$^{3}$/W$^{2}$, and $I_{S}=4.4\times10^{14}$ W/m$^2$. The value obtained for $I_S$ is the same as found by Gu \textit{et al.} In the limit of low intensities $\alpha_{3}$ is equal to the result of He \textit{et al.} \cite{he 2005}.

The cooling time $k(n_{tot})$ is density-dependent. \cite{sun 2005} Carrier cooling times in the range of 30 fs to 1.75 ps have been reported in literature.\cite{sun 2005, yamamoto 1999, takeda 2002, szarko 2005, song 2005, tisdale 2008} By fitting the Saturation and Cooling Model to our experimental results we found
\begin{equation}
k(n_{tot})=k_{0}+r n_{tot},
\end{equation}
with $k_0=0.2$ ps and $r=4\times10^{-39}$ m$^3$s.

It is interesting to compare our results with the 266-nm pump-THz probe results of Hendry \textit{et al.}\cite{hendry 2007} They also found a fast initial decay of the carrier density, followed by a plateau. In their analysis, the decay was attributed to Auger recombination.

\section{\label{z-scan}Z-scan measurement}

In order to test the 800-nm 3PA coefficient obtained in Sec. \ref{Saturation and Cooling Model}, an open-aperture Z-scan measurement was performed. The open-aperture Z-scan has been reported for the first time by Sheik-Bahae \textit{et al.}\cite{sheik-bahae 1990} as a sensitive technique to measure non-linear absorption coefficients.

\begin{figure}
\includegraphics[width=0.5\textwidth]{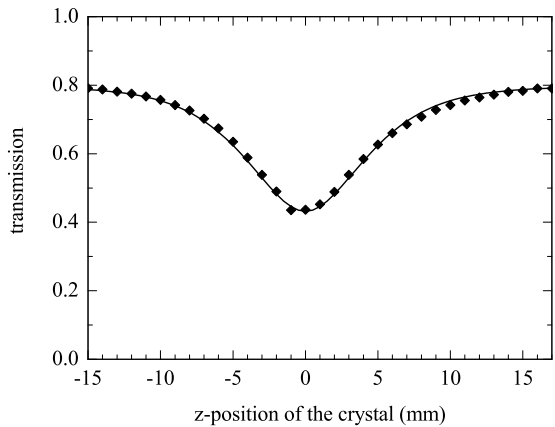}
\caption{Z-scan. The diamonds indicate the measured 800-nm transmission through the 523 $\mu$m thick crystal versus the position of the crystal with respect to the laser focus. The line is the transmission calculated using the intensity-dependent three-photon absorption coefficient found in Sec. \ref{Saturation and Cooling Model}, i.e. Eq. (\ref{3PA saturation}) with $\alpha_{3}^{A}=3\times10^{-27}$ m$^{3}$/W$^{2}$, $\alpha_{3}^{B}=7\times10^{-27}$ m$^{3}$/W$^{2}$, and $I_{S}=4.4\times10^{14}$ W/m$^2$.\label{fig14}}
\end{figure}

For this measurement, the ZnO crystal was moved along the 800-nm beam through the focus (i.e. in the z-direction, hence the name 'Z-scan'). The absolute transmission was measured as a function of the position of the crystal. The results are shown in Fig. \ref{fig14}. For this measurement, the pulse duration was $135 \pm 5$ fs (FWHM), the maximum fluence 355 J/m$^2$, and the maximum intensity $2.2\times10^{15}$ W/m$^2$. The Rayleigh range was measured to be 3.6 mm.

The line in Fig. \ref{fig14} is the transmission through the ZnO crystal, calculated from the intensity-dependent 3PA coefficient $\alpha_{3}[I]$ stated above [Eq. (\ref{3PA saturation})], and also taking into account the 800-nm reflectivity at the front- and backside of the crystal. Evidently, the agreement between the measured and calculated transmission is excellent. This result confirms the value for the 3PA coefficient found in Sec. \ref{Saturation and Cooling Model}. Thus, it is also further evidence for the reliability of the theoretical spectra shown in Figs. \ref{fig9} and \ref{fig11} and the carrier dynamics shown in Fig. \ref{fig13}.

The transmission data of Fig. \ref{fig14} also confirm that the penetration depth of the 800-nm pump pulses is very long compared to the wavelength of the probe. Therefore the 800-nm pump indeed creates a homogeneous carrier density within the penetration depth of the reflected probe, in contrast to a pump with a photon energy above the band gap.

\section{Conclusions}

The ultrafast screening and carrier dynamics in ZnO, including the crossover between the exciton regime and the electron-hole plasma regime, have been studied experimentally and theoretically. Pump-probe reflectivity measurements, taken at four probe wavelengths near the exciton resonance and in a broad range of carrier densities ($10^{22}-10^{26}$ m$^{-3}$), show rapid ($\sim 1$ ps) and strong (up to 30\%) changes in the reflectivity. These effects result from the disappearance of the exciton resonance due to screening. Other processes affecting the optical properties at high densities are band-gap renormalization and band filling.

Our calculations show that the Mott density in ZnO is $1.5\times10^{24}$ m$^{-3}$ at 300 K. This means that phenomena occurring at higher carrier densities in ZnO cannot be related to excitons. This includes lasing in ZnO nanowires and other ZnO nanostructures: if the density is higher than $1.5\times10^{24}$ m$^{-3}$, lasing must be electron-hole plasma lasing. \cite{versteegh to be published 2}

To calculate the optical spectra of highly excited ZnO, we used quantum field theory of a quasi-equilibrium system of electrons and holes that interact via the screened Coulomb potential. We computed the density-dependent spectra of the complex susceptibility, the complex refractive index, the reflectivity, and the absorption coefficient. By comparing the theoretical reflectivity spectra with the pump-probe data, we obtained a description of the carrier dynamics, consisting of 3PA with an intensity-dependent 3PA coefficient, a density-dependent carrier cooling time, and a distinction between surface carriers having a decay time of 0.7 ps and bulk carriers having a decay time of hundreds of picoseconds. The agreement between the theoretical reflectivity based on this model and the experimental results is excellent. Finally, the results of an open-aperture Z-scan confirm the obtained 3PA coefficient.

These results provide strong evidence that this many-body theory well describes screening and band filling in ZnO at high carrier densities and that the band-gap renormalization is approximately given by Eq. (\ref{BGR}). The crossover between the exciton regime and the electron-hole plasma regime, as well as the optical spectra, are faithfully described by this theory. Our results for the density-dependent optical spectra and the ultrafast carrier dynamics are of general importance for ZnO research.

\begin{acknowledgments}
We thank R.E.C. van der Wel for performing the Z-scan measurement, C.R. de Kok and P. Jurrius for technical support, and A.J. van Lange and O.L. Muskens for fruitful discussions.
\end{acknowledgments}

\appendix*
\section{}

For the computation of the optical properties of a semiconductor, an expression for the susceptibility is needed. In this Appendix we derive the susceptibility of an electron-hole gas in a direct-band-gap semiconductor from many-body quantum field theory within random-phase approximation (RPA). In particular, we derive Eq. (\ref{BS susceptibility}) from the statically screened Bethe-Salpeter ladder equation.

\subsection{Polarization and susceptibility}

Consider a direct-band-gap semiconductor crystal, subject to an oscillating external electromagnetic field with angular frequency $\omega$,
\begin{equation}
\mathcal{E}(t)=\mathcal{E}_{0}e^{-i\omega t}.
\end{equation}
In the pump-probe experiment of this paper $\mathcal{E}(t)$ is the field of the probe laser pulse. The electromagnetic field incites a polarization response of the electron-hole gas present in the semiconductor: The polarization oscillates at the same frequency,
\begin{equation}
P(t)=P_{0}e^{-i\omega t}.
\end{equation}
As long as the electric field is not extremely strong, there is a linear relation between $\mathcal{E}_{0}$ and $P_{0}$,
\begin{equation}\label{polarization from electric field}
P_{0}=\chi(\omega)\varepsilon_{0}\mathcal{E}_{0}L^{3}.
\end{equation}
The polarization response is thus described by the complex susceptibility $\chi(\omega)$, where we explicitly indicate the dependence of the susceptibility on the frequency of the electromagnetic field. Our probe laser pulses were weak enough to be in this regime of linear response.

When after the pump pulse the charge carriers have relaxed to Fermi-Dirac distributions at a certain temperature, equilibrium statistical mechanics can be used to describe its properties. The expectation value for $P_{0}$ is given by the relation
\begin{equation}\label{polarization from partition function}
\langle P_{0}\rangle=\frac{1}{\beta}\frac{\partial\ln Z_{gr}}{\partial\mathcal{E}_{0}},
\end{equation}
where $Z_{gr}$ is the grand canonical partition function. Computing this $Z_{gr}$, we use the many-body quantum field theory described in chapter 12 of Ref. \onlinecite{stoof 2009}.

\subsection{Action}

The grand canonical partition function describing the response of the electron and hole fields in the direct-band-gap semiconductor to the external field $\mathcal{E}$ is given by the functional integral
\begin{equation}\label{partition function}
Z_{gr}(\mathcal{E})=\int \mathrm{d}[\phi^{*}]\mathrm{d}[\phi] e^{-S[\phi^{*},\phi,\mathcal{E}^{*},\mathcal{E}]/\hbar},
\end{equation}
where $\phi$ stands for $\phi_{e\uparrow}, \phi_{e\downarrow}, \phi_{h\uparrow}, \phi_{h\downarrow}$, the electron and hole fields with spin up and spin down, respectively, and $S$ is the action, given by
\begin{equation}
\begin{split}
&S[\phi^{*},\phi,\mathcal{E}^{*},\mathcal{E}]\\
&\!\!=\sum_{i,\alpha}\int_{0}^{\hbar\beta}\!\!\!\!\mathrm{d}\tau\!\!\int\!\! \mathrm{d}\mathbf{x}\phi_{i,\alpha}^{*}(\mathbf{x},\tau)\Big\{\hbar\frac{\partial}{\partial\tau}-\frac{\hbar^{2}\nabla^{2}}{2m_{i}}-\mu_{i}\Big\}\phi_{i,\alpha}(\mathbf{x},\tau)\\
&\!\!-\sum_{\alpha}\int_{0}^{\hbar\beta\!\!\!}\mathrm{d}\tau\!\!\int\!\! \mathrm{d}\mathbf{x} \mathrm{d}\mathbf{x'}\phi_{e,\alpha}^{*}(\mathbf{x},\tau)\phi_{h,-\alpha}^{*}(\mathbf{x'},\tau)\\
&\qquad\qquad\qquad\qquad\times V_{s}(\mathbf{x}-\mathbf{x'})\phi_{h,-\alpha}(\mathbf{x'},\tau)\phi_{e,\alpha}(\mathbf{x},\tau)\\
&\!\!-\sum_{\alpha}\int_{0}^{\hbar\beta}\!\!\!\mathrm{d}\tau\!\!\int \!\!\mathrm{d}\mathbf{x}\ d_{cv}\mathcal{E}^{*}(\tau)\phi_{h,-\alpha}(\mathbf{x},\tau)\phi_{e,\alpha}(\mathbf{x},\tau)\\
&\!\!-\sum_{\alpha}\int_{0}^{\hbar\beta}\!\!\!\mathrm{d}\tau\!\!\int\!\! \mathrm{d}\mathbf{x}\ d_{cv}\mathcal{E}(\tau)\phi_{e,\alpha}^{*}(\mathbf{x},\tau)\phi_{h,-\alpha}^{*}(\mathbf{x},\tau).
\end{split}
\end{equation}
Here $\alpha$ stands for $\uparrow$ or $\downarrow$, $i$ again stands for $e$ or $h$, $\tau=it$ is imaginary time, and the Yukawa potential $V_{s}$ is given by Eq. (\ref{yukawa}).

The first term describes the energy of the quasi-free electrons and holes (quasi-free because the interactions between the carriers renormalize the band gap). The second term describes the attractive interaction between electrons and holes with opposite spin. The repulsive interactions and the attraction between electrons and holes with parallel spin are contained in the BGR and therefore not represented here. The third term describes the annihilation of electron-hole pairs by the electric field (stimulated emission). The fourth term describes the creation of electron-hole pairs by the electric field (absorption). We consider only transitions without spin-flip, so that the hole of the electron-hole pair always has a spin opposite to the electron spin.

In the rest of this Appendix, we write $fg(x)$ for the product of functions $f(x)g(x)$. In addition, in order to simplify the theory, we replace the interaction potential $V_{s}(\mathbf{x}-\mathbf{x'})$ by a point interaction
\begin{equation*}
V_{s}(\mathbf{x}-\mathbf{x'})\rightarrow-V_{0}\delta(\mathbf{x}-\mathbf{x'}).
\end{equation*}
In the end, we again replace the point interaction by the screened Coulomb potential.

\subsection{Hubbard-Stratonovich transformation}

Now we introduce two fields, $\Delta_{\uparrow}(\mathbf{x},\tau)$ and $\Delta_{\downarrow}(\mathbf{x},\tau)$, of which the averages are given by
\begin{equation}
\langle\Delta_{\uparrow}(\mathbf{x},\tau)\rangle=V_{0}\langle\phi_{h\downarrow}\phi_{e\uparrow}(\mathbf{x},\tau)\rangle,
\end{equation}
\begin{equation}
\langle\Delta_{\downarrow}(\mathbf{x},\tau)\rangle=V_{0}\langle\phi_{h\uparrow}\phi_{e\downarrow}(\mathbf{x},\tau)\rangle.
\end{equation}

With these two fields we perform a Hubbard-Strato\-novich transformation. The following two identities are inserted into the integrand of the partition function, Eq. (\ref{partition function}):
\begin{equation}
\begin{split}
&1=\exp[\textrm{Tr}[\ln(-V_{0}^{-1}/\hbar)]]\int \mathrm{d}[\Delta_{\alpha}^{*}]\mathrm{d}[\Delta_{\alpha}]\\
&\times\exp\Big\{\frac{1}{\hbar}\int_{0}^{\hbar\beta}\!\!\!\mathrm{d}\tau\!\!\int\!\! \mathrm{d}\mathbf{x}[\Delta_{\alpha}^{*}(\mathbf{x},\tau)-\phi_{e,\alpha}^{*}\phi_{h,-\alpha}^{*}(\mathbf{x},\tau)V_{0}]\\
&\times V_{0}^{-1}[\Delta_{\alpha}(\mathbf{x},\tau)-V_{0}\phi_{h,-\alpha}\phi_{e,\alpha}(\mathbf{x},\tau)]\Big\},
\end{split}
\end{equation}
for $\alpha=\uparrow$ and $\alpha=\downarrow$. This cancels the fourth-order term in the action, so that the fermionic integrals become Gaussian. In the following computation we absorb the factor $\exp[\textrm{Tr}[\ln(-V_{0}^{-1}/\hbar)]]$ into the integration measure.

The resulting action, only quadratically depending on the fermionic fields, is
\begin{equation}
\begin{split}
&S[\Delta^{*},\Delta,\phi^{*},\phi,\mathcal{E}^{*},\mathcal{E}]=-\int_{0}^{\hbar\beta}\!\!\mathrm{d}\tau\!\int \!\mathrm{d}\mathbf{x}\frac{|\Delta_{\uparrow}|^{2}+|\Delta_{\downarrow}|^{2}}{V_{0}}\\
&\!\!-\!\hbar\!\sum_{i,\alpha}\!\!\int_{0}^{\hbar\beta}\!\!\!\!\!\mathrm{d}\tau \mathrm{d}\tau'\!\!\!\int\!\!\! \mathrm{d}\mathbf{x} \mathrm{d}\mathbf{x'}\phi_{i,\alpha}^{*}\!(\mathbf{x},\tau)G_{0i}^{-1}\!(\mathbf{x},\!\tau\!;\mathbf{x'}\!,\!\tau')\phi_{i,\alpha}\!(\mathbf{x'},\tau')\\
&\!\!-\sum_{\alpha}\int_{0}^{\hbar\beta}\!\!\!\mathrm{d}\tau\!\!\int\!\! \mathrm{d}\mathbf{x}[d_{cv}\mathcal{E}^{*}(\tau)-\Delta_{\alpha}^{*}(\mathbf{x},\tau)]\phi_{h,-\alpha}\phi_{e,\alpha}(\mathbf{x},\tau)\\
&\!\!-\sum_{\alpha}\int_{0}^{\hbar\beta}\!\!\!\mathrm{d}\tau\!\!\int\!\! \mathrm{d}\mathbf{x}[d_{cv}\mathcal{E}(\tau)-\Delta_{\alpha}(\mathbf{x},\tau)]\phi_{e,\alpha}^{*}\phi_{h,-\alpha}^{*}(\mathbf{x},\tau),
\end{split}
\end{equation}
where $\Delta$ stands for $\Delta_{\uparrow},\Delta_{\downarrow}$ and where $G_{0e}^{-1}(\mathbf{x},\tau;\mathbf{x'},\tau')$ and $G_{0h}^{-1}(\mathbf{x},\tau;\mathbf{x'},\tau')$ are the inverse noninteracting Green's functions, given by
\begin{equation}\label{inverse Greens functions}
G_{0i}^{-1}\!(\mathbf{x},\!\tau\!;\mathbf{x'}\!,\!\tau')\!=\!-\frac{1}{\hbar}\Big\{\hbar\frac{\partial}{\partial\tau}-\frac{\hbar^{2}\nabla^{2}}{2m_{i}}-\mu_{i}\Big\}\delta(\mathbf{x}-\mathbf{x'})\delta(\tau-\tau').
\end{equation}

\subsection{Integration over the fermion fields}

The integrand of the action can be written in the form of a matrix multiplication as
\begin{equation}
\begin{split}
&S[\Delta^{*},\Delta,\phi^{*},\phi,\mathcal{E}^{*},\mathcal{E}]=-\!\int_{0}^{\hbar\beta}\!\!\!\mathrm{d}\tau\!\!\int\!\! \mathrm{d}\mathbf{x}\frac{|\Delta_{\uparrow}|^{2}+|\Delta_{\downarrow}|^{2}}{V_{0}}\\
&-\hbar\!\int_{0}^{\hbar\beta}\!\!\mathrm{d}\tau \mathrm{d}\tau'\!\int\! \mathrm{d}\mathbf{x} \mathrm{d}\mathbf{x'}\\
&\times\left( \begin{array}{c}
\phi_{e\uparrow}^{*}(\mathbf{x},\tau)\\
\phi_{h\downarrow}(\mathbf{x},\tau)\\
\phi_{e\downarrow}^{*}(\mathbf{x},\tau)\\
\phi_{h\uparrow}(\mathbf{x},\tau)
\end{array} \right)\cdot\mathbf{G}^{-1}(\mathbf{x},\tau;\mathbf{x'},\tau')\cdot\left( \begin{array}{c}
\phi_{e\uparrow}(\mathbf{x'},\tau')\\
\phi_{h\downarrow}^{*}(\mathbf{x'},\tau')\\
\phi_{e\downarrow}(\mathbf{x'},\tau')\\
\phi_{h\uparrow}^{*}(\mathbf{x'},\tau')
\end{array} \right).
\end{split}
\end{equation}

Here, $\mathbf{G}^{-1}$ is the inverse Green's function matrix, which can be expressed in a noninteracting part and a self-energy part
\begin{equation}
\mathbf{G}^{-1}(\mathbf{x},\tau;\mathbf{x'},\tau')=\mathbf{G}^{-1}_{0}(\mathbf{x},\tau;\mathbf{x'},\tau')-\mathbf{\Sigma}(\mathbf{x},\tau;\mathbf{x'},\tau'),
\end{equation}
where the noninteracting part $\mathbf{G}^{-1}_{0}(\mathbf{x},\tau;\mathbf{x'},\tau')$ is given by the matrix
\begin{equation}
\left( \begin{array}{cccc}
\vspace{1mm}\!\!G^{-1}_{0e}(\mathbf{x},\tau;\mathbf{x'},\tau') & \!\!\!\!\!\!\!\!\!\!\!\!\!\!\!\!\!\!\!\!0 & \!\!\!\!\!\!\!\!\!\!\!\!\!\!\!\!\!\!\!\!0 & \!\!\!\!\!\!\!\!\!\!\!\!\!\!\!\!\!\!\!\!0\\
\vspace{1mm}\!\!\!\!0 & \!\!\!\!\!\!\!\!\!\!\!\!\!\!\!\!\!\!\!\!-G^{-1}_{0h}(\mathbf{x}',\tau';\mathbf{x},\tau) & \!\!\!\!\!\!\!\!\!\!\!\!\!\!\!\!\!\!\!\!0 & \!\!\!\!\!\!\!\!\!\!\!\!\!\!\!\!\!\!\!\!0\\
\vspace{1mm}\!\!\!\!0 & \!\!\!\!\!\!\!\!\!\!\!\!\!\!\!\!\!\!\!\!0 & \!\!\!\!\!\!\!\!\!\!\!\!\!\!\!\!\!\!\!\!G^{-1}_{0e}(\mathbf{x},\tau;\mathbf{x'},\tau') & \!\!\!\!\!\!\!\!\!\!\!\!\!\!\!\!\!\!\!\!0\\
\!\!\!\!0 & \!\!\!\!\!\!\!\!\!\!\!\!\!\!\!\!\!\!\!\!0 & \!\!\!\!\!\!\!\!\!\!\!\!\!\!\!\!\!\!\!\!0 & \!\!\!\!\!\!\!\!\!\!\!\!\!\!\!\!\!\!\!\!-G^{-1}_{0h}(\mathbf{x}',\tau';\mathbf{x},\tau)
\end{array} \!\!\right),
\end{equation}
and the selfenergy is
\begin{equation}
\begin{split}
&\mathbf{\Sigma}(\mathbf{x},\tau;\mathbf{x'},\tau')=\frac{1}{\hbar}\delta(\mathbf{x}-\mathbf{x'})\delta(\tau-\tau')\\
&\times\left( \begin{matrix}
\vspace{1mm}\!0 &\!\!\Lambda_{\uparrow}(\mathbf{x},\tau) & \!\!0 & \!\!0\\
\vspace{1mm}\!\Lambda_{\uparrow}^*(\mathbf{x},\tau) & \!\!0 & \!\!0 & \!\!0\\
\vspace{1mm}\!0 & \!\!0 & \!\!0 & \!\!\Lambda_{\downarrow}(\mathbf{x},\tau)\\
\!0 & \!\!0 & \!\!\Lambda_{\downarrow}^*(\mathbf{x},\tau) & \!\!0
\end{matrix}\!\right),
\end{split}
\end{equation}
where $\Lambda_{\alpha}(\mathbf{x},\tau)=\Delta_{\alpha}(\mathbf{x},\tau)-d_{cv}\mathcal{E}(\tau)$.

We can now perform the integration over the fermion fields, using the well-known results for Gaussian integrals. \cite[p. 28]{stoof 2009} The result is the effective action
\begin{equation}
S^{\textrm{eff}}[\Delta^{*}\!\!,\!\Delta,\mathcal{E}^{*}\!\!,\!\mathcal{E}]\!=\!-\!\!\int_{0}^{\hbar\beta}\!\!\!\!\!\mathrm{d}\tau\!\!\!\int\!\!\mathrm{d}\mathbf{x}\frac{|\Delta_{\uparrow}|^{2}\!\!+\!|\Delta_{\downarrow}|^{2}}{V_{0}}
-\hbar \textrm{Tr}[\ln(-\mathbf{G}^{-1}\!)],
\end{equation}
related to the partition function as
\begin{equation}
Z_{gr}\!=\!\!\int \!\!\mathrm{d}[\Delta^{*}]\mathrm{d}[\Delta]e^{-S^{\textrm{eff}}[\Delta^{*},\Delta,\mathcal{E}^{*},\mathcal{E}]/\hbar}.
\end{equation}
The trace is to be taken over space, imaginary time and over the $4\times4$ matrix structure of the Green's function (Nambu space).

\subsection{Power expansion in $\Delta(\mathbf{x},\tau)$}

Now we expand the effective action into powers of $\Delta(\mathbf{x},\tau)$. We write
\begin{equation}
\mathbf{G}^{-1}=\mathbf{G}^{-1}_{0} (\mathbf{1}-\mathbf{G}_{0}\mathbf{\Sigma} )
\end{equation}
and make a Taylor expansion of the logarithm,
\begin{equation}
 \textrm{Tr}[\ln(-\mathbf{G}^{-1})]\!=\! \textrm{Tr}[\ln(-\mathbf{G}^{-1}_{0})]-\sum_{m=1}^{\infty}\frac{1}{m}\textrm{Tr}[(\mathbf{G}_{0}\mathbf{\Sigma})^{m}].
\end{equation}

The first-order term ($m=1$) is 0: There is no contribution in the effective action that is first-order in $\Delta(\mathbf{x},\tau)$. For the second order term ($m=2$) we find
\begin{equation}
\begin{split}
&\frac{\hbar}{2}\textrm{Tr}[(\mathbf{G}_{0}\mathbf{\Sigma})^{2}]=-\frac{1}{\hbar}\sum_{\alpha}\int_{0}^{\hbar\beta}\!\!\mathrm{d}\tau \mathrm{d}\tau'\!\int\! \mathrm{d}\mathbf{x}\mathrm{d}\mathbf{x'}\\
&\times G_{0e}(\mathbf{x},\tau;\mathbf{x}',\tau')\Lambda_{\alpha}(\mathbf{x}',\tau')G_{0h}(\mathbf{x},\tau;\mathbf{x}',\tau')\Lambda^{*}_{\alpha}(\mathbf{x},\tau).
\end{split}
\end{equation}

Now we make the approximation to ignore all terms in the effective action with order $>2$.
Within this approximation the effective action can be written in matrix multiplication form as
\begin{equation}\label{effective action as matrix multiplication}
\begin{split}
&S^{\textrm{eff}}[\Delta^{*}\!\!,\!\Delta,\mathcal{E}^{*}\!\!,\!\mathcal{E}]=-\hbar \textrm{Tr}[\ln(-\mathbf{G}^{-1}_{0})]-\!\int_{0}^{\hbar\beta}\!\!\mathrm{d}\tau \mathrm{d}\tau'\!\int \!\mathrm{d}\mathbf{x}\mathrm{d}\mathbf{x'} \\
&\times\bigg\{\begin{pmatrix}\Delta_{\uparrow}^{*}\vspace{1mm}\\ \Delta_{\downarrow}^{*} \end{pmatrix}\cdot\begin{pmatrix}
\frac{1}{V_{0}}\delta\delta+\frac{1}{\hbar}G_{0e}G_{0h} \hspace{40pt}0\vspace{1mm}\\
0\hspace{40pt}  \frac{1}{V_{0}}\delta\delta+\frac{1}{\hbar}G_{0e}G_{0h} \end{pmatrix}\cdot \begin{pmatrix}\Delta_{\uparrow}\vspace{1mm}\\ \Delta_{\downarrow} \end{pmatrix} \\
&+\begin{pmatrix}\Delta_{\uparrow}^{*}\vspace{1mm}\\ \Delta_{\downarrow}^{*} \end{pmatrix}\cdot \begin{pmatrix} -\frac{1}{\hbar}G_{0e}G_{0h}d_{cv}\mathcal{E}(\tau')\vspace{1mm} \\
-\frac{1}{\hbar}G_{0e}G_{0h}d_{cv}\mathcal{E}(\tau') \end{pmatrix}\\
&+ \begin{pmatrix} -\frac{1}{\hbar}G_{0e}G_{0h}d_{cv}\mathcal{E}(\tau)\vspace{1mm} \\
-\frac{1}{\hbar}G_{0e}G_{0h}d_{cv}\mathcal{E}(\tau) \end{pmatrix}\cdot \begin{pmatrix} \Delta_{\uparrow}\vspace{1mm} \\ \Delta_{\downarrow} \end{pmatrix} \\ &+\frac{2}{\hbar}G_{0e}G_{0h}d_{cv}\mathcal{E}^{*}(\tau)d_{cv}\mathcal{E}(\tau')
\bigg\},
\end{split}
\end{equation}
where $(\Delta_{\uparrow}^{*},\Delta_{\downarrow}^{*})$ is a shorthand for $(\Delta_{\uparrow}^{*}(\mathbf{x},\tau),\Delta_{\downarrow}^{*}(\mathbf{x},\tau))$, $(\Delta_{\uparrow},\Delta_{\downarrow})$ for $(\Delta_{\uparrow}(\mathbf{x'},\tau'),\Delta_{\downarrow}(\mathbf{x'},\tau'))$, $\delta\delta$ for $\delta(\mathbf{x}-\mathbf{x}')\delta(\tau-\tau')$, and $G_{0e}G_{0h}$ for $G_{0e}(\mathbf{x},\tau;\mathbf{x}',\tau')G_{0h}(\mathbf{x},\tau;\mathbf{x}',\tau')$.

\subsection{Bethe-Salpeter ladder equation}

In the normal phase the expectation value for the $\Delta$-fields is zero. It only becomes nonzero if the temperature decreases below a certain critical temperature, far below room temperature. If that happens, condensation of electron-hole Cooper pairs occurs and the system becomes a superfluid. \cite{versteegh to be published 3} Here we describe the optical properties in the normal state. Since the effective action up to this order is quadratic in the $\Delta$-fields, the integration over them can easily be performed: \cite[p. 26]{stoof 2009}
\begin{equation}\label{partition function 2}
Z_{gr}(\mathcal{E})=e^{-S[\mathcal{E}^{*},\mathcal{E}]/\hbar},
\end{equation}
where $S[\mathcal{E}^{*},\mathcal{E}]$ equals, including the factor absorbed into the integration measure,
\begin{equation}
\begin{split}
&S[\mathcal{E}^{*},\mathcal{E}]=S_{BS}[\mathcal{E},\mathcal{E}^{*}]-2\hbar\textrm{Tr}[\ln(-V_{0}^{-1}/\hbar)]\\
&\!-\!\hbar \textrm{Tr}[\ln(-\mathbf{G}^{-1}_{0})]+\!\textrm{Tr}[\ln(-\!\begin{pmatrix}
 \frac{1}{V_{0}}\delta\delta+\frac{1}{\hbar}G_{0e}G_{0h} \hspace{20pt}0\vspace{1mm} \\0\hspace{20pt}  \frac{1}{V_{0}}\delta\delta+\frac{1}{\hbar}G_{0e}G_{0h} \end{pmatrix})].
\end{split}
\end{equation}
Note that the last term was obtained by integration over boson fields, while the third term resulted from integration over fermion fields. This explains the opposite signs.

The term $S_{BS}[\mathcal{E}^{*},\mathcal{E}]$ is the Bethe-Salpeter light-mat\-ter action. Being the sum of all action terms containing $\mathcal{E}(\tau)$, it determines the optical properties. From now on, we concentrate purely on this term. The action $S_{BS}[\mathcal{E}^{*},\mathcal{E}]$ is given by
\begin{equation}\label{light-matter action}
S_{BS}[\mathcal{E}^{*},\mathcal{E}]=S_{MF}[\mathcal{E}^{*},\mathcal{E}]+S_{CA}[\mathcal{E}^{*},\mathcal{E}].
\end{equation}

\begin{figure}
\includegraphics[width=0.35\textwidth]{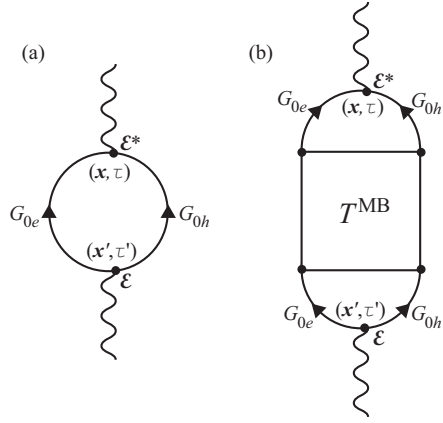}
\caption{Feynman diagrams of the Bethe-Salpeter light-matter action: (a) Mean field, (b) Electron-hole Coulomb attraction.\label{figA1}}
\end{figure}

The term $S_{MF}[\mathcal{E}^{*},\mathcal{E}]$ here is the mean-field action
\begin{equation}\label{mean-field action}
S_{MF}[\mathcal{E}^{*},\mathcal{E}]\!=\!-\!\!\int_{0}^{\hbar\beta}\!\!\!\mathrm{d}\tau \mathrm{d}\tau'\!\!\int \!\!\mathrm{d}\mathbf{x}\mathrm{d}\mathbf{x'}d_{cv}^2\mathcal{E}^{*}(\tau)\frac{2}{\hbar}G_{0e}G_{0h}\mathcal{E}(\tau').
\end{equation}
The mean-field action is the part of the action without $V_{0}$, so without the attractive interaction between electrons and holes with opposite spin. It is like a free-particle term; we call it a `quasi-free-particle term', since the BGR includes the other Coulomb interactions. Figure \ref{figA1}(a) shows the Feynman diagram of the mean-field action.

The Cou\-lomb attraction term, $S_{CA}[\mathcal{E}^{*},\mathcal{E}]$, describes the effect of the attractive interaction between electrons and holes with opposite spin on the optical properties. The integration over the $\Delta$-fields shows that it equals
\begin{equation}\label{Coulomb attraction term}
\begin{split}
&S_{CA}[\mathcal{E}^{*},\mathcal{E}]=\int_{0}^{\hbar\beta}\!\!\mathrm{d}\tau \mathrm{d}\tau' \mathrm{d}\tau''\mathrm{d}\tau'''\!\int\! \mathrm{d}\mathbf{x}\mathrm{d}\mathbf{x'}\mathrm{d}\mathbf{x}''\mathrm{d}\mathbf{x}'''\\
&\times d_{cv}\mathcal{E}^{*}(\tau)\frac{2}{\hbar}G_{0e}(\mathbf{x},\tau;\mathbf{x}'',\tau'')G_{0h}(\mathbf{x},\tau;\mathbf{x}'',\tau'')\\
&\times(\frac{1}{V_{0}}\delta\delta+\frac{1}{\hbar}G_{0e}G_{0h})^{-1}(\mathbf{x}'',\tau'';\mathbf{x}''',\tau''')\\
&\times\frac{1}{\hbar}G_{0e}(\mathbf{x}''',\tau''';\mathbf{x}',\tau')G_{0h}(\mathbf{x}''',\tau''';\mathbf{x}',\tau')d_{cv}\mathcal{E}(\tau').
\end{split}
\end{equation}
Eqs. (\ref{light-matter action}-\ref{Coulomb attraction term}) constitute the Bethe-Salpeter ladder equation.

\begin{figure*}
\includegraphics[width=\textwidth]{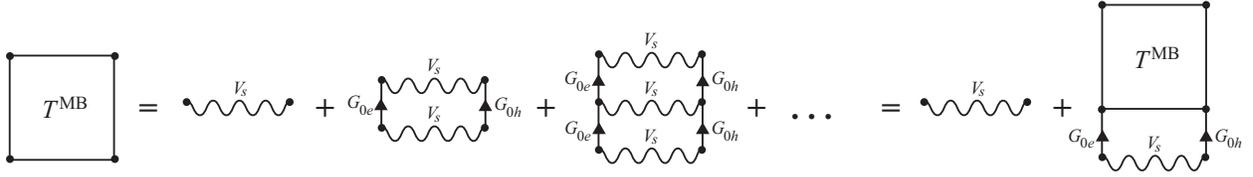}
\caption{Bethe-Salpeter ladder equation: Expansion of the many-body $T$ matrix into ladder diagrams.\label{figA2}}
\end{figure*}

\begin{figure*}
\includegraphics[width=\textwidth]{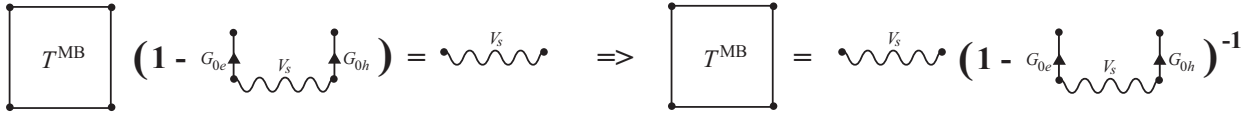}
\caption{Derivation of $S_{CA}[\mathcal{E}^{*},\mathcal{E}]$ [Eq. (\ref{Coulomb attraction term})] via Feynman diagrams.\label{figA3}}
\end{figure*}

Figure \ref{figA1}(b) shows the Feynman diagram of the Cou\-lomb attraction term. The square denotes the many-body $T$ matrix. It consists of the ladder sum of diagrams shown in Fig. \ref{figA2}. The interaction $V_s$ here is the statically screened Cou\-lomb potential (Yukawa potential). By performing a little algebra with the diagrams, as shown in Figs. \ref{figA2} and \ref{figA3}, we obtain a result for the many-body $T$ matrix that can be recognized in Eq. (\ref{Coulomb attraction term}), where a point interaction replaces the Yukawa potential. We see here that the Bethe-Salpeter light-matter action can be derived in two ways: analytically in the way explained above, and alternatively via Feynman diagrams.

\subsection{Fourier transformation}

In order to find the susceptibility as a function of frequency the light-matter action has to be transformed from coordinate space to momentum space and from imaginary time space to imaginary frequency space. We start by transforming the Green's functions.

The inverse Green's functions are defined by Eq. (\ref{inverse Greens functions}). By using the definition of the inverse of a matrix in coordinate and imaginary-time space,
\begin{equation}\label{definition inverse matrix}
\begin{split}
&\delta(\mathbf{x}-\mathbf{x}'')\delta(\tau-\tau')\\
&=\int_{0}^{\hbar\beta}\!\!\mathrm{d}\tau'\!\int \!\mathrm{d}\mathbf{x}'M_{\alpha,\alpha'}(\mathbf{x},\tau;\mathbf{x}',\tau')M_{\alpha',\alpha''}^{-1}(\mathbf{x}',\tau';\mathbf{x}'',\tau''),
\end{split}
\end{equation}
we find for the Green's functions
\begin{equation}\label{Greens functions}
G_{0i}(\mathbf{x},\!\tau;\mathbf{x'}\!,\tau')\!=\!\frac{1}{\hbar \beta L^{3}}\!\sum_{\mathbf{k}}\!\!\!\sum_{n=-\infty}^{\infty}\!\!\frac{-\hbar e^{i\mathbf{k}\cdot(\mathbf{x}-\mathbf{x}')}e^{-i\omega_{n}(\tau-\tau')}}{-i\hbar\omega_{n}+\varepsilon_{\mathbf{k},i}-\mu_{i}},
\end{equation}
where $\varepsilon_{\mathbf{k},i}=\varepsilon_{k,i}=\hbar^{2}k^{2}/(2m_{i})$, and $\omega_{n}$ are the fermionic Matsubara frequencies $\omega_{n}=\pi (2n+1)/(\hbar \beta)$.

For Fourier transforming the screened Coulomb potential we use Eqs. (\ref{Fourier potential}), but for other functions we adopt the conventions
\begin{equation}
\begin{split}
f(\mathbf{x},\tau)=&\frac{1}{\sqrt{L^{3}\hbar\beta}}\sum_{\mathbf{k}}\sum_{n=-\infty}^{\infty}f(\mathbf{k},i\omega_{n})e^{i\mathbf{k}\cdot\mathbf{x}}e^{-i\omega_{n}\tau};\\
f(\mathbf{k},i\omega_{n})=&\frac{1}{\sqrt{L^{3}\hbar\beta}}\int_{0}^{\hbar\beta}\!\!\mathrm{d}\tau\!\int \!\mathrm{d}\mathbf{x} f(\mathbf{x},\tau)e^{-i\mathbf{k}\cdot\mathbf{x}}e^{i\omega_{n}\tau},
\end{split}
\end{equation}
Further,
\begin{equation}\label{delta functions}
\sum_{\mathbf{k}}\!e^{i\mathbf{k}\cdot(\mathbf{x}-\mathbf{x}')}\!=L^{3}\delta (\mathbf{x}-\mathbf{x}'), \text{ and}
\int\!\frac{\mathrm{d} \mathbf{x}}{L^{3}}e^{-i(\mathbf{k}-\mathbf{k}')\cdot\mathbf{x}}\!=\delta_{\mathbf{k},\mathbf{k}'}.
\end{equation}

The Fourier transform of Eq. (\ref{Greens functions}) is given by
\begin{equation}
G_{0i}(\mathbf{k},i\omega_{n};\mathbf{k}',i\omega_{n'})=\delta_{\mathbf{k},\mathbf{k}'}\delta_{n,n'}\frac{-\hbar}{-i\hbar\omega_{n}+\varepsilon_{\mathbf{k},i}-\mu_{i}}
\end{equation}
or, in a shorter notation,
\begin{equation}\label{Fourier Greens functions}
G_{0i}(\mathbf{k},i\omega_{n})=\frac{-\hbar}{-i\hbar\omega_{n}+\varepsilon_{\mathbf{k},i}-\mu_{i}}.
\end{equation}

Being a photon field, the oscillating electromagnetic field is expressed in quantum field theory in bosonic Matsubara frequencies $\omega_{f}=2\pi f/(\hbar\beta)$, and in imaginary time $\tau=i t$,
\begin{equation}\label{photon field}
\mathcal{E}(\tau)=\frac{1}{\sqrt{\hbar\beta}}\mathcal{E}_{f}e^{-i\omega_{f} \tau},\:\:\text{and}\quad
\mathcal{E}^{*}(\tau)=\frac{1}{\sqrt{\hbar\beta}}\mathcal{E}^{*}_{f}e^{i\omega_{f} \tau},
\end{equation}
where $\mathcal{E}_{f}=\sqrt{\hbar\beta}\ \mathcal{E}_{0}$. The relation between the Matsubara frequency and the real frequency is given by the Wick rotation $i\hbar\omega_{f}\rightarrow\hbar\omega-E_G+\mu_{e}+\mu_{h}$. Because we work here with complex electric fields instead of real fields, an extra factor $1/2$ has to be included into the terms quadratic in $\mathcal{E}$.

\subsection{Mean-field susceptibility}

\begin{figure}
\includegraphics[width=0.45\textwidth]{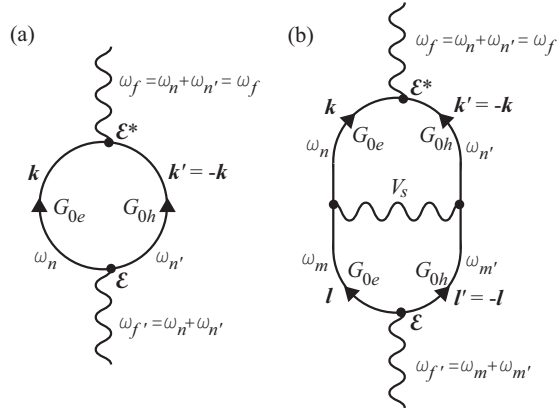}
\caption{Feynman diagrams of the Bethe-Salpeter light-matter action in momentum and Matsubara frequency space: \mbox{(a) Mean field,} (b) Simplest diagram of the electron-hole Coulomb attraction.\label{figA4}}
\end{figure}

Inserting Eqs. (\ref{Greens functions}), (\ref{Fourier Greens functions}), and (\ref{photon field}) into Eq. (\ref{mean-field action}), and elaborating the integrals over coordinate space and imaginary time, we obtain for the mean-field action
\begin{equation}\label{Fourier mean-field action}
\begin{split}
&S_{MF}[\mathcal{E}^{*},\mathcal{E}]=-\frac{1}{\hbar^{2}\beta}\sum_{\mathbf{k},\mathbf{k}'}\sum_{n,n'}d_{cv}^2\mathcal{E}^{*}_{f}G_{0e}(\mathbf{k},i\omega_{n})\\
&\times G_{0h}(\mathbf{k}',i\omega_{n'})\mathcal{E}_{f'}\delta_{-\mathbf{k},\mathbf{k}'}\delta_{\mathbf{k},-\mathbf{k}'}\delta_{-\omega_{f},-\omega_{n}-\omega_{n'}}\delta_{\omega_{f'},\omega_{n}+\omega_{n'}}\\
&=-\frac{1}{\hbar^{2}\beta}d_{cv}^{2}|\mathcal{E}_{f}|^{2}\sum_{\mathbf{k}}\sum_{n}G_{0e}(\mathbf{k},i\omega_{n})G_{0h}(-\mathbf{k},i\omega_{f-n-1})\\
&=-\frac{1}{\hbar^{2}\beta}d_{cv}^{2}|\mathcal{E}_{f}|^{2}\sum_{\mathbf{k}}\sum_{n}\frac{-\hbar}{-i\hbar\omega_{n}+\varepsilon_{\mathbf{k},e}-\mu_{e}}\\
&\times\frac{-\hbar}{-i\hbar\omega_{f-n-1}+\varepsilon_{-\mathbf{k},h}-\mu_{h}},
\end{split}
\end{equation}
where we used Eq. (\ref{delta functions}). The Kronecker deltas describe the conservation of momentum and energy. Figure \ref{figA4}(a) shows the corresponding Feynman diagram. The momentum of a created electron is equal and opposite to the momentum of the created hole. The relatively very small momentum of the absorbed and created photons is neglected in this derivation. The third and fourth Kronecker deltas imply $f=n+n'+1=f'$, meaning that if an electron-hole pair created by a photon recombines again, the emitted photon has the same frequency as the absorbed photon.

We split the fraction and perform the sum over $n$ by contour integration, \cite[p. 140]{stoof 2009}
\begin{equation}
\begin{split}
&S_{MF}[\mathcal{E}^{*},\mathcal{E}]\!=\!\frac{d_{cv}^{2}|\mathcal{E}_{f}|^{2}}{\hbar^{2}\beta}\lim_{\eta\downarrow0}\sum_{\mathbf{k},n}\frac{\hbar}{-i\hbar\omega_{f}\!+\!\varepsilon_{\mathbf{k},e}\!+\!\varepsilon_{\mathbf{k},h}\!-\!\mu_{e}\!-\!\mu_{h}}\\
&\times\Big(\frac{-\hbar e^{i\omega_{n}\eta}}{-i\hbar\omega_{n}+\varepsilon_{\mathbf{k},e}-\mu_{e}}+\frac{-\hbar e^{i\omega_{-f+n+1}\eta}}{-i\hbar\omega_{f-n-1}+\varepsilon_{\mathbf{k},h}-\mu_{h}}\Big)\\
&=d_{cv}^{2}\hbar\beta\mathcal{E}_{0}^{2}\sum_{\mathbf{k}}\frac{1-f_{\mathbf{k},e}-f_{\mathbf{k},h}}{i\hbar\omega_{f}-\varepsilon_{\mathbf{k},e}-\varepsilon_{\mathbf{k},h}+\mu_{e}+\mu_{h}},
\end{split}
\end{equation}
where $f_{\mathbf{k},e}=f_{k,e}$ and $f_{\mathbf{k},h}=f_{k,h}$ are the Fermi-Dirac electron and hole distribution functions.

Wick rotating to real frequencies and including a fre\-quency-dependent damping factor,
\begin{equation}
i\hbar\omega_{f}\rightarrow\hbar(\omega+i\gamma(\omega))-E_G+\mu_{e}+\mu_{h},
\end{equation}
we obtain for the mean-field action
\begin{equation}
S_{MF}[\mathcal{E}^{*},\mathcal{E}]=-d_{cv}\hbar\beta\mathcal{E}_{0}^{2}\sum_{\mathbf{k}}\chi_{\mathbf{k}}^{0}(\omega),
\end{equation}
with
\begin{equation}\label{Achik0}
\chi_{\mathbf{k}}^{0}(\omega)=\chi_{k}^{0}(\omega)=-d_{cv}\frac{1-f_{k,e}-f_{k,h}}{\hbar(\omega+i\gamma(\omega))\!-\!\varepsilon_{k,e}\!-\!\varepsilon_{k,h}\!-\!E_G}.
\end{equation}

This mean-field function can be used to compute the optical properties within mean-field theory, i.e., by only taking into account the mean-field action. The mean-field contribution to the partition function is
\begin{equation}
Z_{gr}^{MF}(\mathcal{E})=e^{-S_{MF}[\mathcal{E}^{*},\mathcal{E}]/\hbar}=\exp[d_{cv}\beta\mathcal{E}_{0}^{2}\sum_{\mathbf{k}}\chi_{\mathbf{k}}^{0}(\omega)]
\end{equation}
which, using Eq. (\ref{polarization from partition function}), gives a mean-field polarization of
\begin{equation}
<P_{0}>_{MF}=2d_{cv}\mathcal{E}_{0}\sum_{\mathbf{k}}\chi_{\mathbf{k}}^{0}(\omega),
\end{equation}
so, according to Eq. (\ref{polarization from electric field}), the mean-field susceptibility equals
\begin{equation}\label{mean field susceptibility}
\chi_{MF}(\omega)=\frac{2d_{cv}}{\varepsilon_{0}L^{3}}\sum_{\mathbf{k}}\chi_{\mathbf{k}}^{0}(\omega).
\end{equation}

\subsection{RPA susceptibility}

In order to obtain the susceptibility from the full Bethe-Salpeter ladder equation (i.e. within RPA), we need to transform also the Coulomb attraction term [Eq. (\ref{Coulomb attraction term})] into momentum and Matsubara frequency space.

First the factor \mbox{$(\frac{1}{V_{0}}\delta\delta+\frac{1}{\hbar}G_{0e}G_{0h})^{-1}$} has to be expanded. From Eq. (\ref{definition inverse matrix}) we see that
\begin{equation}\label{ladder sum}
\begin{split}
&\!(\frac{1}{V_{0}}\delta\delta+\frac{1}{\hbar}G_{0e}G_{0h})^{-1}(\mathbf{x}',\tau';\mathbf{x}'',\tau'')\\
&\!=V_{0}\delta(\mathbf{x}'-\mathbf{x}'')\delta(\tau'-\tau'')\\
&-\frac{V_{0}^{2}}{\hbar}G_{0e}G_{0h}(\mathbf{x}',\tau';\mathbf{x}'',\tau'')\\
&\!+\!\!\frac{V_{0}^{3}}{\hbar^{2}}\!\!\int_{0}^{\hbar\beta}\!\!\!\!\!\mathrm{d}\tau'''\!\!\!\!\int \!\!\mathrm{d}\mathbf{x}'''G_{0e}G_{0h}(\mathbf{x}'\!,\!\tau'\!;\!\mathbf{x}'''\!,\!\tau''')G_{0e}G_{0h}(\mathbf{x}'''\!,\!\tau'''\!;\!\mathbf{x}''\!,\!\tau'')\\
&\!-\!\frac{V_{0}^{4}}{\hbar^{3}}\!\int_{0}^{\hbar\beta}\!\!\mathrm{d}\tau'''\mathrm{d}\tau''''\!\int \!\!\mathrm{d}\mathbf{x}'''\mathrm{d}\mathbf{x}''''G_{0e}G_{0h}(\mathbf{x}',\tau';\mathbf{x}''',\tau''')\\
&\!\times\! G_{0e}G_{0h}(\mathbf{x}'''\!,\!\tau'''\!;\!\mathbf{x}''''\!,\!\tau'''')G_{0e}G_{0h}(\mathbf{x}''''\!,\!\tau''''\!;\!\mathbf{x}''\!,\!\tau'')\\
&+\ldots
\end{split}
\end{equation}
In this expansion the ladder sum of Fig. \ref{figA2} can be recognized.

By inserting Eq. (\ref{ladder sum}) into Eq. (\ref{Coulomb attraction term}), and by using again Eqs. (\ref{Greens functions}), (\ref{Fourier Greens functions}), (\ref{photon field}), and (\ref{delta functions}), we obtain for the Coulomb attraction term
\begin{equation}
\begin{split}
&S_{CA}[\mathcal{E}^{*},\mathcal{E}]=\frac{1}{2\hbar\beta}d_{cv}^{2}|\mathcal{E}_{f}|^{2}\\
&\times\Big\{\frac{2V_{0}}{\hbar^{3}\beta L^{3}}\sum_{\mathbf{k},\mathbf{l}}\sum_{n,m}G_{0e}(\mathbf{k},i\omega_{n})G_{0h}(-\mathbf{k},i\omega_{f-n-1})\\
&\qquad\times G_{0e}(\mathbf{l},i\omega_{m})G_{0h}(-\mathbf{l},i\omega_{f-m-1})\\
&-\frac{2V_{0}^{2}}{\hbar^{5}\beta^{2}L^{6}}\sum_{\mathbf{k},\mathbf{l},\mathbf{m}}\sum_{n,m,p}G_{0e}(\mathbf{k},i\omega_{n})G_{0h}(-\mathbf{k},i\omega_{f-n-1})\\
&\qquad\times G_{0e}(\mathbf{l},i\omega_{m})G_{0h}(-\mathbf{l},i\omega_{f-m-1})\\
&\qquad\times G_{0e}(\mathbf{m},i\omega_{p})G_{0h}(-\mathbf{m},i\omega_{f-p-1})\\
&+\ldots\Big\}.
\end{split}
\end{equation}

Figure \ref{figA4}(b) shows the Feynman diagram corresponding to the first of these terms. Again the principles of conservation of momentum and energy follow from the theory. Every next term is equal to the previous one, multiplied by
\begin{equation*}
\frac{-V_{0}}{\hbar^{2}\beta L^{3}}\sum_{\mathbf{k}}\sum_{n}G_{0e}(\mathbf{k},i\omega_{n})G_{0h}(-\mathbf{k},i\omega_{f-n-1}).
\end{equation*}
Apart from a prefactor, this factor is equal to the mean-field action [see Eq. (\ref{Fourier mean-field action})]. It can therefore be written as
\begin{equation*}
\frac{-V_{0}}{d_{cv}L^{3}}\sum_{\mathbf{k}}\chi_{\mathbf{k}}^{0}(\omega).
\end{equation*}
It directly follows that the Bethe-Salpeter light-matter action, including the whole Bethe-Salpeter ladder, equals
\begin{equation}
S_{BS}[\mathcal{E}^{*},\mathcal{E}]=-d_{cv}\hbar\beta\mathcal{E}_{0}^{2}\sum_{\mathbf{k}}\chi_{\mathbf{k}}(\omega),
\end{equation}
where
\begin{equation}
\chi_{\mathbf{k}}(\omega)=\chi_{\mathbf{k}}^{0}(\omega)\Big(1-\frac{V_{0}}{d_{cv}L^{3}}\sum_{\mathbf{k}'}\chi_{\mathbf{k}'}(\omega)\Big).
\end{equation}

Finally we have to replace the point interaction by the screened Coulomb potential,
\begin{equation*}
-V_{0}\delta(\mathbf{x}-\mathbf{x'})\rightarrow V_{s}(\mathbf{x}-\mathbf{x'}).
\end{equation*}
The expression for the susceptibility demands a potential in momentum space. From Eqs. (\ref{Fourier potential}) and (\ref{delta functions}) it follows that in momentum space $V_{0}$ has to be replaced by $-V_{s,|\mathbf{k}-\mathbf{k'}|}$, so
\begin{equation}
\chi_{\mathbf{k}}(\omega)=\chi_{k}^{0}(\omega)\Big(1+\frac{1}{d_{cv}L^{3}}\sum_{\mathbf{k}'}V_{s,|\mathbf{k}-\mathbf{k'}|}\chi_{\mathbf{k}'}(\omega)\Big).
\end{equation}

Using Eqs. (\ref{partition function 2}), (\ref{polarization from partition function}), and (\ref{polarization from electric field}), one then easily finds that the susceptibility of the electron-hole gas is given by
\begin{equation}
\chi_{BS}(\omega)=\frac{2d_{cv}}{\varepsilon_{0}L^{3}}\sum_{\mathbf{k}}\chi_{\mathbf{k}}(\omega).
\end{equation}

\end{document}